%% file: paper.tex
\documentclass[fleqn,usenatbib]{mnras}

\usepackage{newtxtext,newtxmath}
\usepackage[T1]{fontenc}
\usepackage[utf8]{inputenc}
\usepackage{ae,aecompl}
\usepackage{graphicx}
\usepackage{amsmath}	
\usepackage{amssymb}	
\usepackage{booktabs}

\usepackage{url}
\usepackage{soul}
\usepackage{pgf,tikz}
\usepackage{mathrsfs}
\usetikzlibrary{fit,positioning}
\usetikzlibrary{calc,matrix,arrows,decorations.pathmorphing}
\usepackage{adjustbox}
\usepackage[colorinlistoftodos,prependcaption,textsize=tiny]{todonotes}
\usepackage{xcolor}
\usepackage{enumerate}

\newcommand{\angstrom}{\mbox{\normalfont\AA}}
\newcommand{\OHb}{$\log$([OIII]/H$\beta$)}
\newcommand{\NHa}{$\log$([NII]/H$\alpha$)}
\newcommand{\EWH}{$\log$ EW(H${\alpha}$)}
\newcommand{\EWNII}{$\log$ EW([NII])}

\newcommand*{\orcid}{\includegraphics[scale = 0.0525]{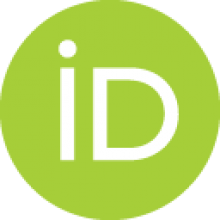}}
\title[The UV upturn fraction within UV bright RSGs]{UV bright red-sequence galaxies: how do UV upturn systems evolve in redshift and stellar mass?}

\author[Dantas, M.~L.~L. et al.]
{\href{https://orcid.org/0000-0002-1178-8169}{M.~L.~L.~Dantas\orcid}$^{1,2}$\thanks{E-mail: mlldantas@protonmail.com},
\href{https://orcid.org/0000-0003-1846-4826}{P.~R.~T.~Coelho\orcid}$^{1}$, \href{https://orcid.org/0000-0001-7207-4584}{R.~S.~de Souza\orcid}$^{3}$, \href{https://orcid.org/0000-0003-2374-366X}{T.~S.~Gonçalves\orcid}$^{4}$\\
$^{1}$Instituto de Astronomia, Geofísica e Ciências Atmosféricas, Universidade de São Paulo, R. do Matão 1226, 05508-090, São Paulo, \\ Brazil\\
$^{2}$Departamento de Física Teórica, Universidad Autónoma de Madrid, 28049, Madrid, Spain\\
$^{3}$Department of Physics \& Astronomy, University of North Carolina at Chapel Hill, NC 27599-3255, USA \\
$^{4}$Observatório do Valongo, Universidade Federal of Rio de Janeiro, Ladeira Pedro Antônio 43, Rio de Janeiro, RJ 20080-090, Brazil}

\date{Accepted XXX. Received YYY; in original form ZZZ}
\pubyear{2019}

\begin{document}
\label{firstpage}
\pagerange{\pageref{firstpage}--\pageref{lastpage}}
\maketitle

\begin{abstract}
The so-called ultraviolet (UV) upturn of elliptical galaxies is a phenomenon characterised by the up-rise of their fluxes in bluer wavelengths, typically in the 1,200--2,500 \angstrom~ range. This work aims at estimating the rate of occurrence of the UV upturn over the entire red-sequence population of galaxies that show significant UV emission. This assessment is made considering it as function of three parameters: redshift, stellar mass, and -- what may seem counter-intuitive at first -- emission-line classification.
We built a multiwavelength spectro-photometric catalogue from the Galaxy Mass Assembly survey, together with aperture-matched data from Galaxy Evolution Explorer Medium-Depth Imaging Survey (MIS) and Sloan Digital Sky Survey, covering the redshift range between 0.06 and 0.40. From this sample, we analyse the UV emission among UV bright galaxies, by selecting those that occupy the red-sequence \emph{locus} in the (NUV-$r$) $\times$ (FUV-NUV) chart; then, we stratify the sample by their emission-line classes. To that end, we make use of emission-line diagnostic diagrams, focusing the analysis in retired/passive lineless galaxies. Then, a Bayesian logistic model was built to simultaneously deal with the effects of all galaxy properties (including emission-line classification or lack thereof). 
The main results show that retired/passive systems host an up-rise in the fraction of UV upturn for redshifts between 0.06 and 0.25, followed by an in-fall up to 0.35. Additionally, we show that the fraction of UV upturn hosts rises with increasing stellar mass.
\end{abstract}


\begin{keywords}
galaxies: general, evolution, elliptical and lenticular, cD -- galaxies: stellar content -- ultraviolet: general -- methods: statistical
\end{keywords}

\section{Introduction} \label{sec:intro}

Galaxies are building blocks of the large scale structure of the Universe; they are a composite of stars, gas, dust, and dark matter, yet the bulk of their electromagnetic radiation comes from their stars and dust. Therefore, exploring the role of their stellar populations is imperative in order to build a comprehensive picture of galaxy evolution. These systems can be split into two major classes \emph{grosso modo}: \emph{early} and \emph{late}-type galaxies -- both terms were coined in the notorious work by \citet{Hubble1926}. 
With respect to early-type systems, their apparent simplicity should not be overlooked; in fact, they are far from `simple': they undergo a series of mergers and other types of dynamic interactions \citep[e.g.][]{Barnes1988, Springel2001, DeLucia2007, Naab2014, Schawinski2014}; these mergers can be classified as wet or dry, playing a critical role in the subsequent steps of their evolution \citep[e.g.][]{SB2009b}. Nonetheless, galaxies, in order to mature into early-types, are subject to a series of feedback processes such as supernovae and active galactic nuclei \citep*[AGN, e.g.][]{Springel2005}; conversely, the formation and evolution of the most massive objects known  -- monstrous passive early-types -- is still not completely understood \citep[e.g.][]{Stoppacher2019}. Ultimately, they can even present a wide range of morphologies, although being dominated by ellipticals and lenticulars \citep[e.g.][]{SB2009a}. 

Given this scenario, the ultraviolet (UV) range of the electromagnetic spectrum provides insights regarding the stellar content of galaxies \citep[e.g. better estimation of star formation rates, more precise metallicity measurements, see][to enumerate a few]{Salim2007, Vazdekis2016}. This emission is associated with hot components, specially young hot stars and is, therefore, a reliable indicator of star formation activity \citep{Kennicutt1998, GildePaz2007, Salim2007, MadauDickinson2014}. However, \citet{CodeWelch1979} reported, for the first time, a considerably high UV emission in early-type galaxies using a pioneer UV satellite at the time: the Orbiting Astronomical Observatory \citep[OAO,][]{Code69}. This emission, detected in galaxies dominated by old stellar populations, is the the so-called \emph{UV upturn of elliptical galaxies}, a pronounced flux increase at bluer wavelengths, around 1,200--2,500 \angstrom \, present in some ellipticals \citep[see e.g.][for a review]{Dorman1993, ferguson_davidsen93, OConnell1999}. This phenomenon was once considered a puzzle, as formerly it was believed that early-type galaxies would be solely composed by cold old stars; this finding levelled up the complexity regarding the overall comprehension of such galaxies then. Later studies have shown that not only these objects can have some residual star formation activity \citep[e.g. ][]{Yi2005, Pipino2009, Salim2010, Bettoni2014, Davis2015, Haines2015M, Stasinska2015, Sheen2016, Vazdekis2016, LopezCorredoira2018}, but also that some post-main-sequence stellar evolutionary phases are also efficient UV emitters \citep[e.g. post-asymptotic giant branch -- post-AGB --, blue horizontal branch -- HB -- stars, binary systems in interaction, and so on,][and references therein]{Greggio1990, Brown1999, OConnell1999, Greggio1999, deharveng2002, Brown2004, Yoon2004, Maraston2005, Han2007, Coelho2009, Donahue2010, Loubser2011, Hernandez-Perez2014}.

In the local Universe, evidence that the UV upturn is generated by extreme horizontal branch (EHB) stars has been accumulating \citep{brown+98, brown+00, BC03, Maraston2005, peng_nagai09, Donahue2010, Loubser2011, Schombert2016}. The cores of these helium-burning stars are covered by a very thin hydrogen layer, exposing them and their higher inner temperatures; hence, boosting the UV emission of quiescent galaxies. Additionally, recent studies indicate that helium enhanced populations could explain both the UV properties of globular clusters and the UV upturn phenomenon \citep[e.g.][]{lee+05, piotto+07, Schiavon2012, Chung2017, Peacock2018}.

Nevertheless, it is yet unclear why some -- but not all -- galaxies dominated by old stellar populations develop such strong UV emission, and how the UV upturn is linked to other features of the host galaxy. It seems to be more common among luminous, high velocity dispersion galaxies \citep{burstein+88}, but previous works seem to indicate that the dependency with mass is weak \citep{Yi2011}. \citet{Rich2005} found no correlation between UV rising flux and other parameters sensitive to the global metallicity of such systems. Conversely, it seems that the strength of the upturn correlates positively with optical metallicity indicators \citep[e.g.][]{Yoon2004, Chung2017, Ali2018a}, which is expected in case the emission is primarily due to helium enhanced populations. But against the expectations of helium sedimentation theories \citep[e.g.][which predict a stronger UV flux in brightest cluster galaxies -- BCGs -- when compared to other ellipticals]{peng_nagai09}, there has been no strong evidence of a dependency of the UV phenomenon with their host environment \citep[for instance, ][]{Yi2011, Ali2019}. Yet, by using absorption lines, \citet{LeCras2016} revealed that young and old stellar populations co-exist, which has triggered a renewed interest in this subject.

The presence of AGN could also contribute to the UV flux \citep[e.g.][]{Cid2000, Heinis2016, Padovani2017} in red-sequence galaxies (RSGs); those are often classified as low-ionisation nuclear emission-line region (LINERs): their ionisation mechanisms are historically attributed to AGN activity -- which has been strongly questioned, for many of such galaxies seem to be ionised by hot low-mass evolved (post-main sequence) stars \citep[AKA HOLMES,][]{CidFernandes2011, Sing2013, Belfiore2016, Padovani2017, Herpich2018}. On the other hand, it seems that AGN activity contributes only to a tiny part of the far-UV emission in systems with very bright nuclei \citep[e.g. M87,][]{Ohl1998}.

The evolution of the \emph{strength} of the UV upturn with redshift ($z$) has been tackled by numerous articles \citep[e.g.][]{brown+98,brown+00,Brown2004,Rich2005,Ree2007}. While \citet{Rich2005} have analysed a potential trend in the (FUV-$r$) colour in $z$ bins, no correlation was found. \citet{Brown2004} used a different approach, by analysing the evolved stellar populations in a handful of galaxies and comparing differences in UV-optical colours in terms of lookback time, suggesting an evolution in the UV upturn, up to $z=0.6$. Also, \citet{Ree2007} examined the (FUV-V) colour of nearby elliptical galaxies and those in clusters up to $z=0.2$, as well as other two clusters at $z=0.3$ and $z=0.5$; this study indicated a feeble evolution of this colour. Recently \citeauthor{Ali2018a} published a series of papers revisiting the UV upturn discussion \citep{Ali2018a, Ali2018b, Ali2018c, Ali2019} as well as \citet{Boissier2018}. To explore the evolution of the UV upturn, \citet{Ali2018c} selected data from four galaxy clusters (one at $z\approx0.31$, two at $z\approx0.55$, and one at $z\approx0.68$), retrieved from the Hubble Space Telescope (HST) Legacy  Archive\footnote{\url{https://hla.stsci.edu/}}; they concluded that the incidence of the phenomenon rises up until $z\approx0.55$ and then it strongly declines, suggesting that, in fact, the UV upturn depends on redshift.

The purpose of the present work is to analyse the proportion of UV upturn galaxies against the larger group of UV bright red-sequence systems, and how it changes at a given $z$, as well as their dependency with stellar mass ($\log M_{\star}$ in units of solar masses -- $M_{\odot}$ -- which is omitted throughout the text for simplicity). To that end, we selected a sample of red-sequence galaxies bright enough in the UV to be detected in both Galaxy Evolution Explorer bands; consequently, this sample was classified according to the paradigm proposed by \cite{Yi2011}. Then, using only red-sequence systems (NUV-$r\geq 5.4$) we model the incidence of UV upturn among them, by making use of a customised Bayesian logistic regression. To deal with completeness biases, the $\log M_{\star}$ is directly embedded in the analysis (see Sec. \ref{sec:methodology}). Finally, as a complementary step, the sample is stratified by their emission-line classes (whenever present, through the BPT, \citealt{Baldwin81}, and specially the WHAN diagrams, \citealt{Cid2010, CidFernandes2011}). With the combination of both classifications (UV and emission-lines), the sample is then refined allowing us to deliberate on the probable surrogates of the UV upturn.  

This paper is structured as follows: the sample is described in Sec. \ref{sec:catalogue}; an overview of the sample is presented in Sec. \ref{sec:overview_rsg}; the statistical model is described in Sec. \ref{sec:methodology}; the analysis is portrayed in Sec. \ref{sec:analysis}; the discussion is developed in Sec. \ref{sec:discussion}; and, finally, in Sec. \ref{sec:conclusions} the conclusions are presented. Extra material can be found in the Appendix \ref{sec:app_bpt}. The adopted cosmology is the standard $\Lambda$-Cold Dark Matter ($\Lambda$CDM) throughout this paper, following the procedure adopted by the GAMA team\footnote{We refer the reader to the GAMA schema browser: \url{http://www.gama-survey.org/dr3/schema/dmu.php?id=9} -- paragraph `DISTANCES AND COSMOLOGY'.}. Hence the following parameters were used: $\left\lbrace H_{0}, ~\Omega_{M}, ~\Omega_{\Lambda} \right\rbrace = \left\lbrace 70 {\rm km}~{\rm s}^{-1}~{\rm Mpc}^{-1}, {\rm 0.3, ~0.7}  \right\rbrace$.

\section{The Sample} \label{sec:catalogue}

This Section describes the selection criteria used to build our sample, as well as the treatment applied.

\subsection{Data Selection} \label{subsec:data_selec}

We built a multiwavelength catalogue using selected galaxies from the Galaxy and Mass Assembly Data Release 3 survey \citep[GAMA-DR3,][]{Liske2015, Baldry2018} with photometric observations retrieved from: the Galaxy Evolution Explorer Data Release 6/plus 7 \citep[GALEX,][for a more recent review]{Martin2005,Bianchi2014} Medium-depth Imaging Survey (MIS) in the UV, and (model magnitudes) from the Sloan Digital Sky Survey Data Release 7 \citep[SDSS-DR7, ][]{Abazajian2009} in the optical; all sources were aperture-matched (from the table in the UV \texttt{ApMatchedCat}). 

The sample was retrieved by matching the following main catalogues in GAMA-DR3 data management units (DMUs), \texttt{ApMachtedCat}, \texttt{InputCatA} with \texttt{GalexMain}. Also, spurious measurements were considered and treated, according to the flags listed in order:

\begin{itemize}
    \item all matches were made using the catalogue identification, \texttt{CATAID}, from GAMA's database;
    
    \item only galaxies with measurements in all five bands -- two from GALEX and five from SDSS -- were considered. Missing data in the UV photometry (i.e. -9999.0) were also removed;

    \item objects that were knowingly galaxies were selected (hence, \texttt{TYPE}=3 provided by \texttt{InputCatA});
    
    \item objects which had more than one match both for GALEX and GAMA were discarded (\texttt{NMATCHUV}=1 and \texttt{NMATCHOPT}=1);
    
    \item UV observations with potential problems were also rejected, i.e \texttt{NUVFLAG}=0 and \texttt{FUVFLAG}=0;
    
    \item all $z$'s used are heliocentric and estimated by spectral measurements made with the AAOmega/2dF, instruments used for the GAMA survey \citep[for details on spectroscopic measurements, we refer the reader to][]{Hopkins2013};
    
    \item constraints on the $z$ range were `organically' applied: $z_{min}$=0.06 and $z_{max}$=0.4 (minimum and maximum redshift, respectively -- we refer the reader to Sec. \ref{sec:overview_rsg}); 
    
    \item restrictions on the quality of $z$ measurements ware taken into account by adopting \texttt{NQ}$>$2 and \texttt{PROB}$>$0.8, as recommended by \cite{Baldry2018}.
\end{itemize}

Regarding the spectral signal, which is important for emission-line measurements, no constraint was imposed on the signal-to-noise ratio ($S/N$), given that it could have an impact on the number of objects in the final sample, as discussed in \citet{Cid2010}. In Sec. \ref{subsec:finalsample}, the description of the $S/N$ in the final sample is made, but we anticipate that only $\approx 0.4$ per cent of it is comprised by objects with $S/N<3$ -- the standard cut for emission-line detection \citep[e.g.][]{kauffmann2003}.

The $\log M_{\star}$ inference is described in details by \citet{Taylor2011}. The emission-line measurements were retrieved from table \texttt{GaussFitComplex} inside \texttt{SpecLineSFR} DMU in GAMA's database. Finally, we briefly discuss internal extinction in Sec. \ref{subsec:app_kcorr_extinct}, by making use of \texttt{MagPhys} DMU which contains output parameters -- such as dust mass ($\log M_{\rm{dust}}$)\footnote{In units of $M_{\odot}$ herein omitted for simplicity.} -- from SED fitting performed by the GAMA team using the MagPhys code by \citet{DaCunha2008}.

\subsection{Data Treatment} \label{subsec:data_treat}

The sample was corrected for the foreground extinction using the maps of \cite{Schlegel1998}, the extinction law described by \cite{Fitzpatrick99} and implemented using the \textsc{python} package \textsc{pyneb} \citep{Luridiana2015}; magnitude offsets in the visible were also taken into account \citep[as described in][]{Doi2010}. Parameters dependent on the adopted cosmology, such as luminosity distance, were estimated using the \textsc{python} package \textsc{astropy} \citep{Astropy2013}. 

K-corrections were adopted as provided by the GAMA team, from the  \texttt{kCorrections} DMU -- with $z=0.0$ and based on SDSS-DR7 model magnitudes\footnote{{\url{http://www.gama-survey.org/dr3/schema/dmu.php?id=7}}}. These corrections were estimated using the code \textsc{k\_correct} \citep{Blanton2007}, making use of \citet{BC03} stellar population templates and \citet{Chabrier2003} initial mass function (IMF). For further details, we refer the reader to Fig. \ref{fig:kcorr}, which displays the distributions of k-corrections (SDSS $r$-band and GALEX FUV and NUV bands) for all UV categories according to \citet{Yi2011}.

Thence, our `primary' sample is comprised of 14,331 objects. Further details are presented in Secs. \ref{subsec:uv_charact} and \ref{subsec:finalsample}; the final sample is presented in Sec. \ref{subsec:finalsample}.

\subsection{UV characterisation} \label{subsec:uv_charact}

\begin{figure}
\includegraphics[width=\linewidth]{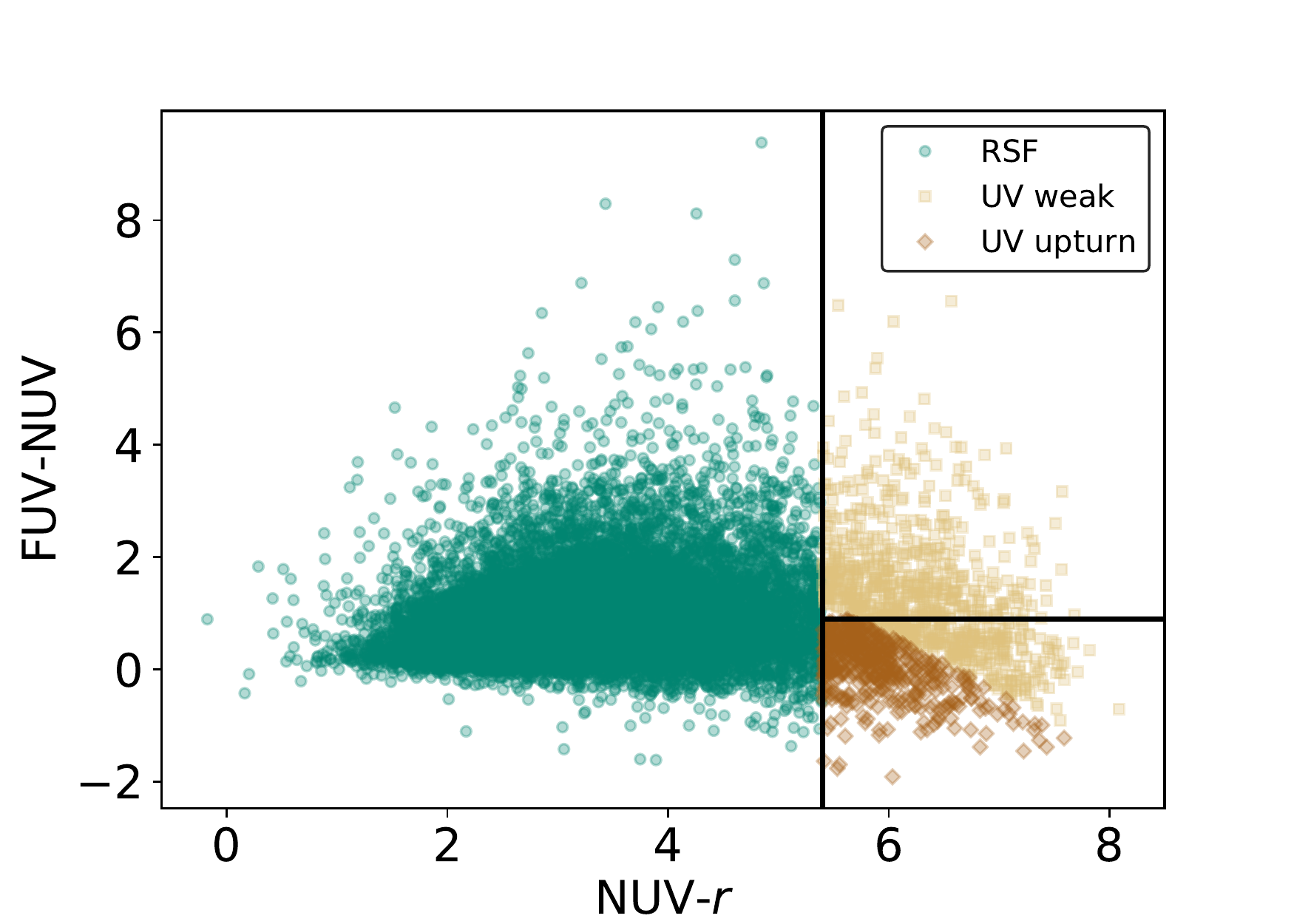}
\caption{UV-optical colour-colour diagnostic diagram for the sample with the proposed classes by \citet{Yi2011}. The green round markers represent galaxies classified as `residual star-formation' (RSF); the light orange squares, the UV weak; and the strong orange diamond-shaped markers, the UV upturn. Systems with (NUV-$r$)>5.4 are our UV-bright RSGs (in both light and dark orange).}
\label{fig:colourcolour_yi}
\end{figure}

The primary sample is depicted in Fig. \ref{fig:colourcolour_yi} in the shape of a colour-colour diagram, including the criteria construed by \citet{Yi2011}. According to the authors, to better restrain potential carriers of UV upturn, one must consider the following criteria: (NUV-$r$) $>$ 5.4, (FUV-NUV) $<$ 0.9, and (FUV-$r$) $<$ 6.6. The first criterion attempts to mitigate the contamination of young stellar populations; in other words, systems with meaningful star-formation are rejected; thence, objects with (NUV-$r$) $>$ 5.4 are the ones we refer to as `UV bright' RSGs. The second criterion measures the rising slope for lower wavelengths, and the third selects objects with substantial FUV flux; hence splitting the UV bright sample into two groups: UV weak and upturn.

\begin{figure}
\includegraphics[width=\linewidth]{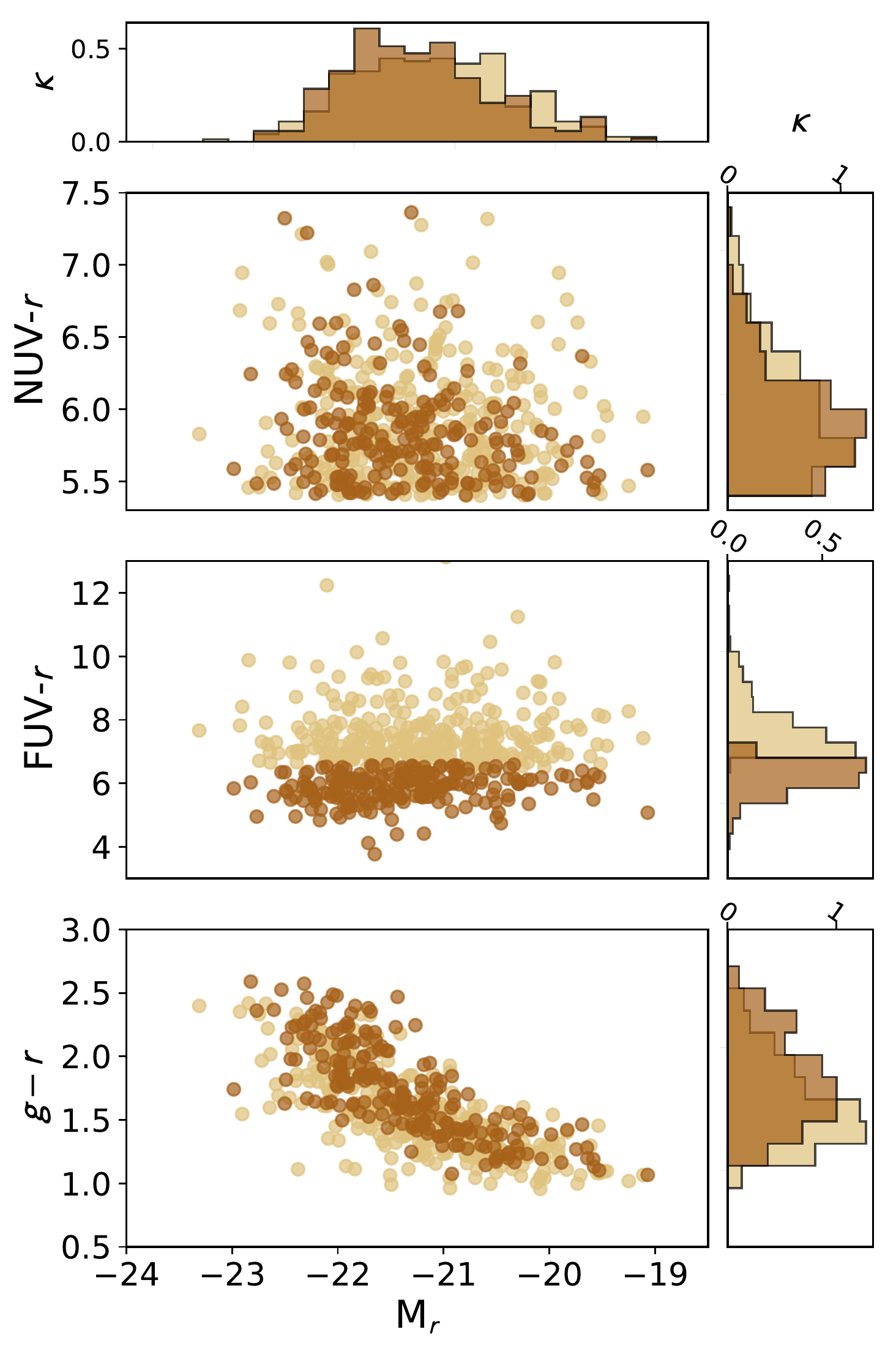}
\caption{Three colour-magnitude charts composed by UV and optical bands are displayed above; all of them are in terms of absolute magnitude in the $r$-band ($M_r$). The top, middle, and bottom panels show, respectively, the diagrams in terms of the following colours: (NUV-$r$), (FUV-$r$), and ($g-r$). Their respective normalised (by area) distributions ($\kappa$) are also displayed in the adjacent histograms.}
\label{fig:color_mag_multi}
\end{figure}

In addition, Fig. \ref{fig:color_mag_multi} features three colour-magnitude diagrams for the UV weak and upturn objects of the sample. The top and middle panel exhibit UV-optical colours \emph{versus} absolute magnitude in SDSS $r$-band ($M_r$), showing the restrictions proposed by \citet{Yi2011}; whereas the bottom panel depicts ($g-r$) \emph{versus} $M_r$, with both UV weak and upturn blended, as expected, since both occupy the red-sequence \emph{locus} \citep{Strateva2001}. Yet, UV upturn systems are slightly redder. The respective distributions of these parameters are displayed on the adjacent histograms.

\subsection{The final sample} \label{subsec:finalsample}

\begin{figure}
    \centering
    \includegraphics[width=\linewidth]{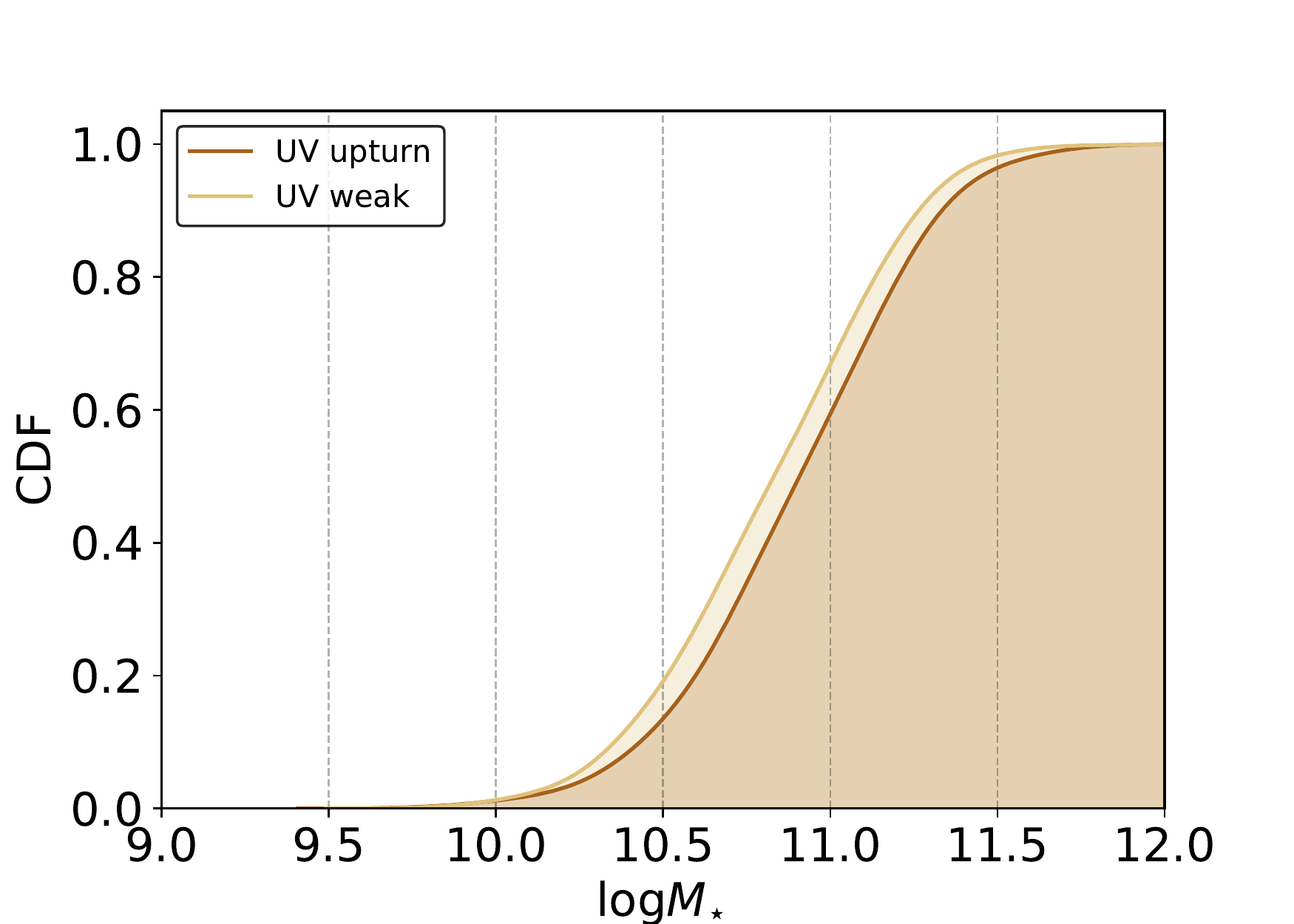}
    \caption{Cumulative distribution function (CDF) of $\log M_{\star}$ for the UV weak and upturn sub-samples herein. The y-axis displays the total amount of galaxies of our sample, being 0 equivalent to 0 per cent and 1 to 100 per cent. There is a small $\log M_{\star}$ shift between the UV weak and upturn CDFs; the latter is composed by slightly more massive systems.}
    \label{fig:mass_cumulative}
\end{figure}

Fig. \ref{fig:mass_cumulative} displays the cumulative function distribution (CDF) for the $\log M_{\star}$ of our sample: only about 20 per cent of galaxies have $\log M_{\star} \leq 10.5$; between 10.5 and 11, the CDF has a whopping jump to around 60 per cent; finally, it nearly reaches 100 per cent for systems with $\log M_{\star} \approx 11.5$. The final sample herewith is comprised of 506 RSGs, being 296 UV weak and 210 upturn. Of these, only 2 objects have $S/N<3$ (or about 0.4 per cent); therefore, the impact in the final results is negligible. Both are UV weak have no detectable emission-line, hence are tagged as unclassified.

The internal extinction of the galaxies in our sample has not been taken into account, therefore, it is likely that some contamination by reddened green-valley systems is present \citep[for further reading on the extinction degeneracy, see][and references therein]{Worthey1994, deMeulenaer2013, Sodre2013}.For further details, we refer the reader to Appendix \ref{subsec:app_kcorr_extinct}.

\section{Overview of the RSG sample} \label{sec:overview_rsg}

The sample described in Sec. \ref{sec:catalogue} is herewith further explored. Fig. \ref{fig:bptwhan_all} illustrates that the UV classes as defined by \citet{Yi2011} are represented in all areas of the emission-line diagrams. For stratified charts given their UV class, we refer the reader to Fig. \ref{fig:bptwhan_split}.

\subsection{UV bright RSGs and their emission-lines} \label{subsec:emlines_charact}

As previously discussed, UV bright RSGs can arise due to different stellar (e.g. residual star-formation, post-main-sequence stellar phases) or non-stellar (e.g. AGN) sources. Therefore, by exploring their emission-line features, it is possible to sketch a general picture of their different potential origins.

\begin{figure} 
\includegraphics[width=\linewidth]{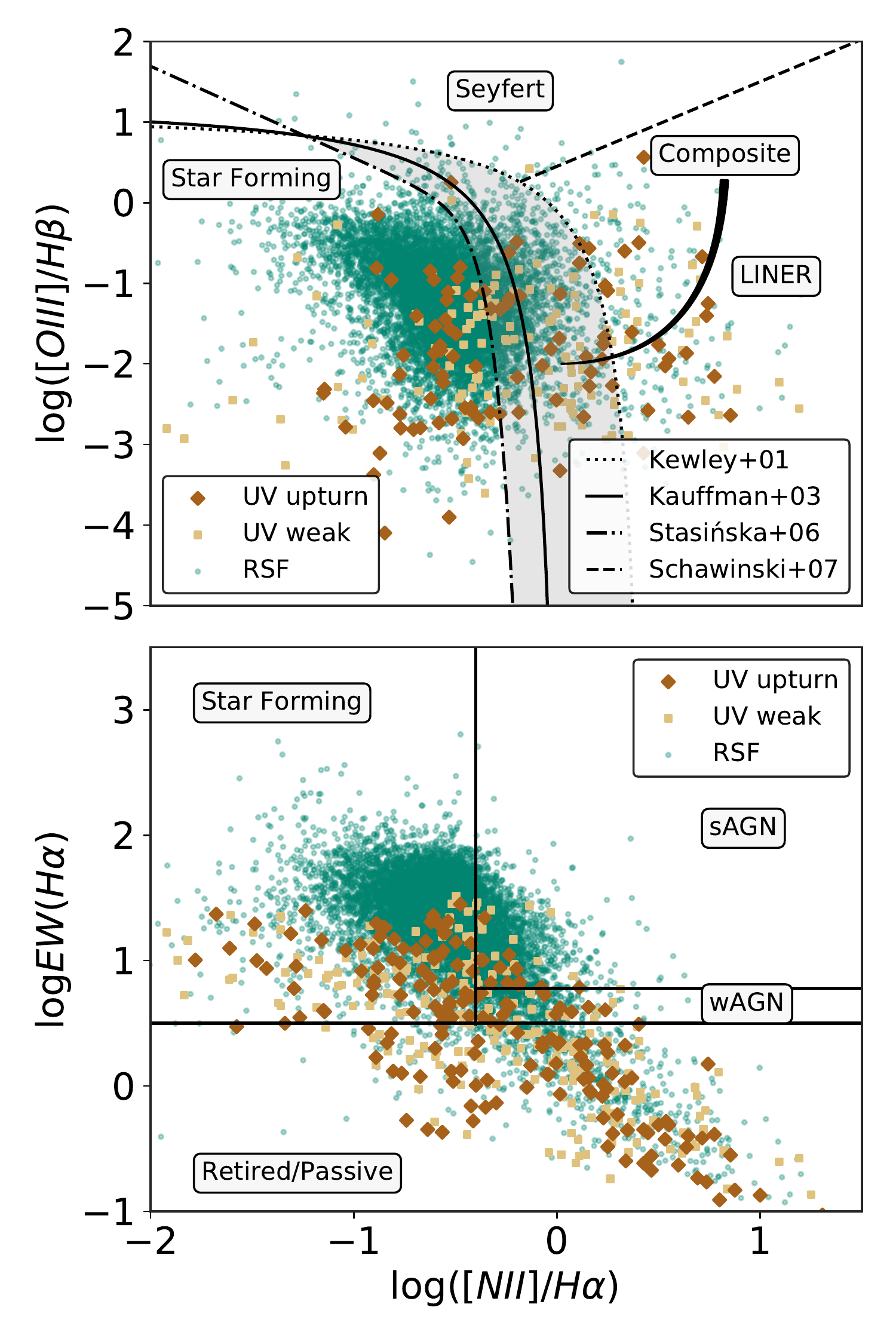}
\caption{The galaxies in our primary sample with detectable emission lines are shown in BPT \citep[top panel,][]{Baldwin81} and WHAN \citep[bottom panel,][]{CidFernandes2011} diagrams, colour-coded by their UV classes: green, light orange, and dark orange represent, in order, RSF, UV weak, and UV upturn. On the BPT diagram (top panel) the division described by \citet{kewley2001} is represented by dotted line; the one by \citet{kauffmann2003} by the solid line; the one by \citet{Stasinska2006} by the dot-dashed line; and that of \citet{Schawinski2007} is represented by the dashed-line. On the WHAN diagram, on the bottom panel, displays the classifications described by \citet{CidFernandes2011}, namely star-forming (SF), strong and weak AGN (sAGN, wAGN), and retired/passive (R/P) galaxies.}
\label{fig:bptwhan_all}
\end{figure}

To isolate the effects of galaxy properties, we make use of emission-line diagnostic diagrams. Henceforth, we make use of the BPT \citep{Baldwin81} and WHAN diagrams \citep{CidFernandes2011}. Fig. \ref{fig:bptwhan_all} displays the aforementioned charts, colour-coded as described in Fig. \ref{fig:colourcolour_yi}. The first shows \NHa \, per \OHb, whereas the latter displays \NHa \, per \EWH. The top panel of Fig. \ref{fig:bptwhan_all} shows the BPT diagram with its notable classes given by the theoretical extreme starburst line by \citet{kewley2001}, the empirical line by \citet{kauffmann2003}, and the hybrid proposition by \citet{Stasinska2006}. Furthermore, the AGN subgroups (Seyferts and LINERs) are also shown, according to the criteria described by \citet{Schawinski2007}. The bottom panel of Fig. \ref{fig:bptwhan_all} displays the WHAN diagram with its classes, according to the description by \citet{CidFernandes2011}, which in brief segragates galaxies into 5 groups: star-forming (SF), strong and weak AGN (sAGN and wAGN), retired galaxies (RG), and passive (lineless) galaxies, as follows:

\begin{itemize}
    \item \NHa$<$-0.4 and \EWH$>$3\angstrom: SF;
    \item \NHa$>$-0.4 and \EWH$>$6\angstrom: sAGN;
    \item \NHa$>$-0.4 and 3\angstrom$\leq$\EWH $\leq$6\angstrom: wAGN;
    \item \EWH$<$3\angstrom: RG;
    \item \EWH \, and \EWNII$<$0.5\angstrom: passive.
\end{itemize}

Differently from the BPT chart, WHAN enables us to spot the retired and some of the passive population, which is suitable for this investigation. Additionally, a number of objects cannot be detected in neither of them, mostly due to the simple fact that our sample is focused on RSGs, hence including a sizeable amount of lineless galaxies \citep{Herpich2018}; nevertheless, this does not guarantee that all of these objects are, in fact, quiescent, for they can be unclassified due to other unknown issues (e.g. low $S/N$) -- these systems are tagged as `unclassified'. \citet{CidFernandes2011} intuitively tag lineless galaxies that cannot be displayed in the WHAN chart as `undetected'; therefore, the reader must keep in mind that our unclassified galaxies are made of the `undetected' lineless passive systems plus those with missing measurements, as previously described.

It follows that, for the primary sample, 11,647 objects appear in the BPT, whereas 13,050 can be identified in the WHAN chart. Thus, for the purposes or the analysis, we only make use of the WHAN diagram; however, we provide ann extended analysis featuring the BPT in the Appendix \ref{sec:app_bpt}. All RSGs in the final sample, including those tagged as `unclassified', are further discussed in Secs. \ref{sec:analysis} and \ref{sec:discussion}.

\begin{table} 
\caption{The following table displays the number of galaxies of each UV class in the WHAN diagram; it includes all the 13,050 with the required parameters successfully measured for the WHAN plus 1,281 remaining systems with no measurable emission-lines (unclassified depicted as `unc.'). The last two rows, in bold, consist the final sample analysed in this paper.}
\label{table:whan_uv}
\begin{center}
\begin{tabular}{lrrrrrr}
\multicolumn{7}{c}{\textbf{WHAN classification}} \\
\hline
\hline
\textbf{UV class} & SF & sAGN   & wAGN  & R/P  & unc. & \textbf{total}\\
\hline
RSF & 9,253  & 2,156  & 557   & 669  & 1,190 & 13,825\\ 
\textbf{weak} & \textbf{78} & \textbf{19} & \textbf{19} & \textbf{118} & \textbf{62} & \textbf{296}\\ 
\textbf{upturn} & \textbf{68} & \textbf{9} & \textbf{17} & \textbf{87} & \textbf{29} & \textbf{210}\\ 
\end{tabular}
\end{center}
\end{table}

This emission-line analysis puts in evidence that broadband photometry alone is not sufficient to select galaxies hosting classic UV upturn. 

\subsection{UV excess as a function of redshift} \label{subsec:poc}

\begin{figure}
\includegraphics[width=\linewidth]{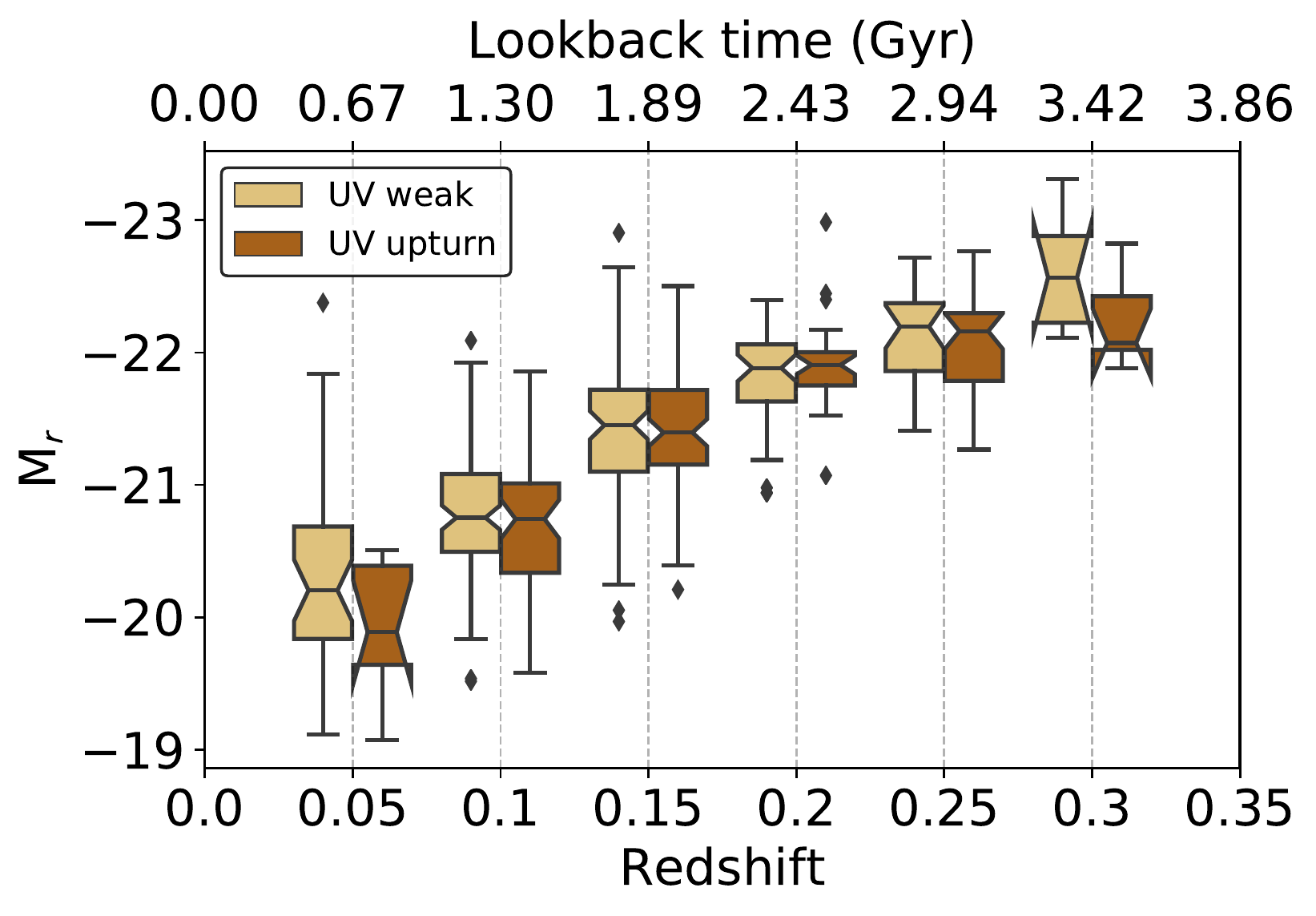}
\caption{Boxplots displaying the distribution of absolute magnitude in the \textit{r}-band ($M_r$) given a set of $z$ bins for the RSGs in our sample. UV weak and upturn are respectively displayed by light and dark shades of orange. The coloured regions of the boxplots display the interquartile range; the bars indicate nearly 3$\sigma$ of the respective distributions; the fliers depict outliers; and the notch encodes the 96 per cent confidence interval around the median $M_r$ displayed by the horizontal line. For further details, we refer the reader to \citet{Tukey1977}.}
\label{fig:boxplots_mag} 
\end{figure}

The distributions of $M_r$ -- Fig. \ref{fig:boxplots_mag} -- as a function of $z$ (bins of 0.05) for both UV weak and upturn groups (i.e. NUV-$r$>5.4) -- are shown as notched  boxplots (the upper x-axis depicts the corresponding lookback time, according to the cosmological parameters described in Sec. \ref{sec:intro}). In each notched boxplot, per $z$ bin and UV group, the notch encodes the 96 per cent confidence interval around the median $M_r$ displayed by the horizontal line. The range of the box depicts the 25 and 75 percentiles of $M_r$. 

\begin{figure}
\includegraphics[width=\linewidth]{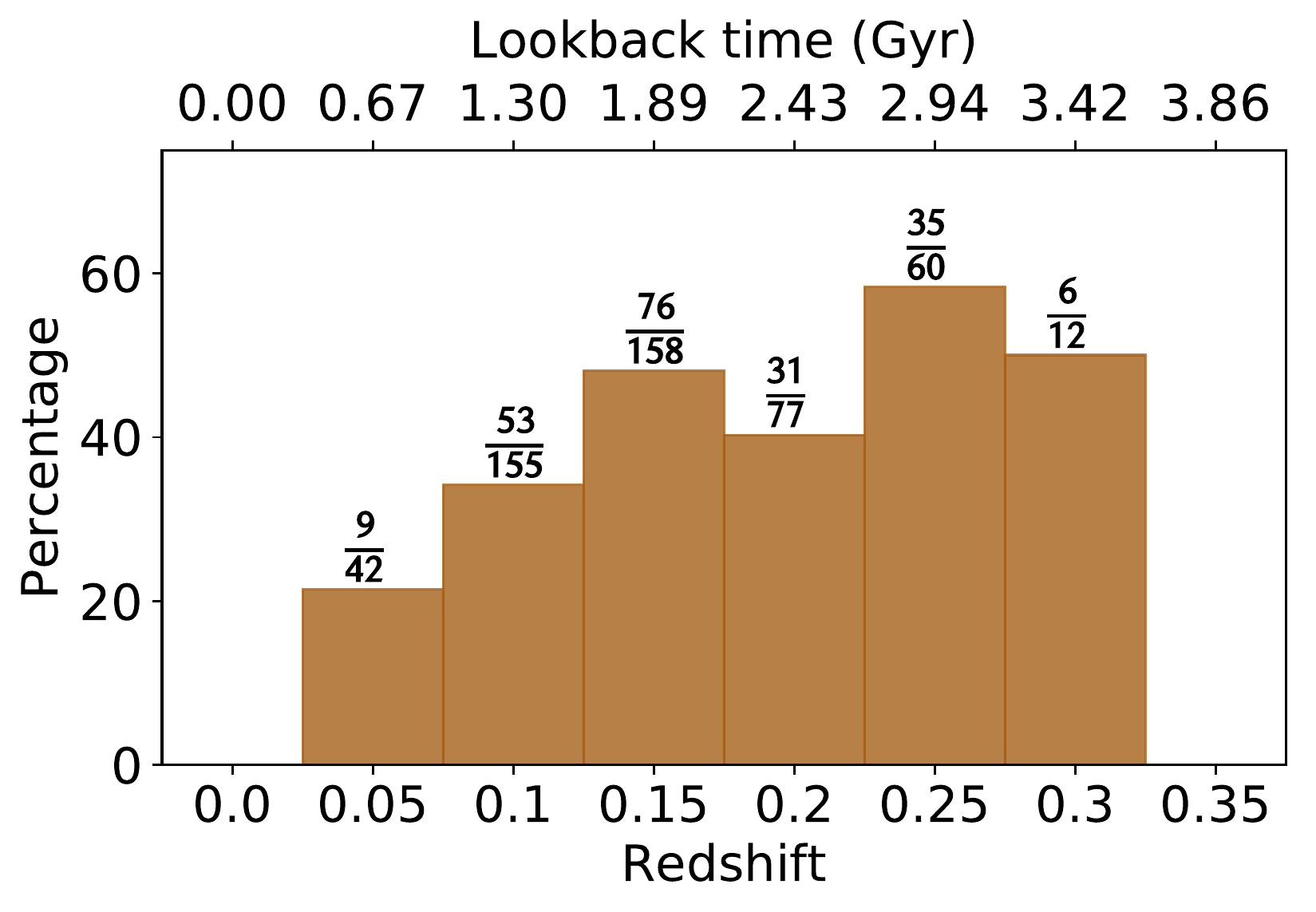}
\caption{Bar chart displaying the percentage of galaxies that present UV upturn in terms of the entire RSGs population (hence, the sum of the UV weak and upturn objects). For each $z$ bin, the numbers corresponding to the \emph{de facto} sum of RSGs and UV upturn bearers, are shown in form of fractions on the top of each bar. The corresponding lookback time is shown in the secondary upper x-axis.}
\label{fig:barplot_uvpercentage}
\end{figure}

Both subgroups of RSGs change monotonically throughout the bins of $z$, being the distribution of the UV upturn group systematically lower in $M_r$ (brighter objects) than the UV weak counterpart. Additionally, the distribution becomes tighter with increasing $z$, that is due to small numbers of objects after $z\approx0.35$. This limit marks the threshold $z$ of the sample. Details on the numbers can be seen further, in Fig. \ref{fig:barplot_uvpercentage}.

To evaluate whether the occurrence of the UV upturn phenomenon changes within $z$, we projected a barplot -- Fig. \ref{fig:barplot_uvpercentage} -- containing the fraction of UV upturn hosts over the sum of the entire RSG population of the primary sample. It is clear that the number of objects decrease rapidly with increasing $z$; the last bin already shows 6 UV upturn objects over 12, which is the total RSGs in such bin. Furthermore, the bins are chosen arbitrarily, being manipulated at will, which can lead to different results as bins change. All in all, this evidence alone is not robust to prove any evolution with $z$, but it is provocative enough to entice a deeper investigation.

\begin{figure*}
\includegraphics[width=\linewidth]{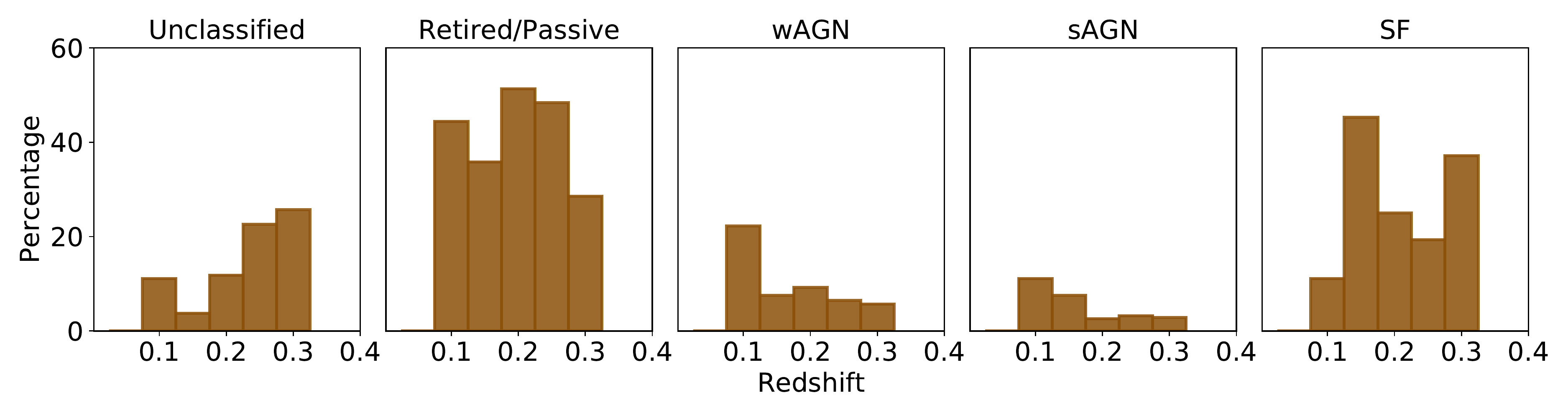}
\caption{Bar charts analogous to Fig. \ref{fig:barplot_uvpercentage}, but stratified by their WHAN classes. The fractions on top of each bar were omitted, as well as the lookback time, for a clearer view of the distributions.}
\label{fig:barplot_uv_split}
\end{figure*}

Fig. \ref{fig:barplot_uv_split} shows the same barplots dividing the sample in WHAN class. To check how this trend may appear given the different emission-line classes, the entire population of UV upturn objects were separated in sub-samples, according to their WHAN class, which is displayed in Fig. \ref{fig:barplot_uv_split}. The distributions change significantly from class to class, but for the unclassified objects, the trend remains. For a robust analysis, we employed a Bayesian model, which is described in Sec. \ref{sec:methodology}.

\section{Statistical Model}\label{sec:methodology}

To model the presence or absence of UV upturn, we employed a Bayesian logistic regression \citep[see][for a review]{Hilbe2017}. Logistic regression has been previously used in astronomy, for example, to probe the likelihood of star-forming activity in primordial galaxies \citep{deSouza2015AeC}, or to model the environmental effects in the presence/absence of AGN \citep{deSouza2016}. It generally aims to model binomial (binary) data. A binomial distribution describes a sequence of independent experiments (trials) each of which has only two possible outcomes $\{0,1\} $. In the specific case of interest here, one can think of the presence of UV upturn as binary data, which is either $1$ for those that  present the phenomenon or $0$ otherwise. We then built two logistic models; the first that simultaneously account for the dependency of the UV upturn on $z$ and $\log M_{\star}$, while the second also considers emission-line classes. The model is portrayed in graphical form at Fig. \ref{fig:matrix}, and reads as follows: each of the $i-th$  galaxies in the dataset, has its probability to manifest UV upturn described as a Bernoulli process, whose probability of success relates to $\log{M_{\star}}$ and $z$ through a logit link function (to ensure the probabilities will fall between 0 and 1) and a second-degree polynomial relation $\eta_{i[k]}$, where the index $k$ encodes the emission-line classification. 

\begin{figure}
\input{model.tikz}
\caption{A matrix representation of the Bayesian Bernoulli model expressing the hierarchy of dependencies between the UV upturn likelihood $\left(p \equiv f_{\rm upturn}\right)$, the stellar mass ($\log M_{\star}$), and redshift ($z$) a data set of galaxies indexed by the subscript \emph{i}. Their respective emission line classifications are indexed by \emph{k}.}
\label{fig:matrix}
\end{figure}

We evaluate our model using a Hamiltonian Monte-Carlo (HMC) sampler, for which we use the domain specific language \textsc{stan} via their \textsc{python} implementation available in the package \textsc{pystan} \citep{Riddell2018}. We initiate four Hamiltonian Markov chains by starting the HMC \citep[e.g.][]{Brooks2011,Carpenter2017} sampler at different initial values. The initial burn-in phases were set to 5,000 steps followed by 15,000 iteration steps for the model considering only $\log M_{\star}$ and $z$; and 3,000 steps followed by 7,000 iteration steps for the model considering $\log M_{\star}$, $z$, and WHAN classes. Those were sufficient to guarantee the convergence of each chain via the so-called Gelman-Rubin statistics \citep{gelman1992}. We assume weakly informative Normal priors for the coefficients $\beta_{jk}$, setting a zero mean Normal prior with standard deviation of 10. 

\section{Results}\label{sec:analysis}

We make use of the entire UV bright RSG sample retrieved from the GAMA-DR3 database, in a continuous range of $z$. The proportion of UV upturn galaxies was then estimated using a logistic regression in order to accurately deal with the phenomenon. Also, by using the fraction of UV upturn objects over the entire UV bright RSGs and embedding the $\log M_{\star}$ into the model, incompleteness's biases such as the Malmquist \citep{Malmquist1922, Sandage2000} are minimised. Nonetheless, we present complementary material by making use of a volume-limited sub-sample in Appendix \ref{sec:app_vollim}, as well as the corresponding results and discussion.

The analysis is split in two complementary parts: the first aims at estimating the probability -- or, in other words, fraction -- of a RSG of hosting the UV upturn emission, according to \citet{Yi2011}; the second part considers the exact same features as the first, except that it also takes into account emission-line classes -- including unclassified/quiescent objects -- by making use of the WHAN diagram. These two parts are presented in Secs. \ref{subsec:results01} and \ref{subsec:results02} respectively.

\subsection{Results for stellar mass and redshift} 
\label{subsec:results01}

The results for the logistic regression considering only $\log M_{\star}$ and $z$ are briefly presented in this Section. All figures display 50 and 95 per cent credible intervals.

\begin{figure*}
    \centering
    \includegraphics[width=\linewidth]{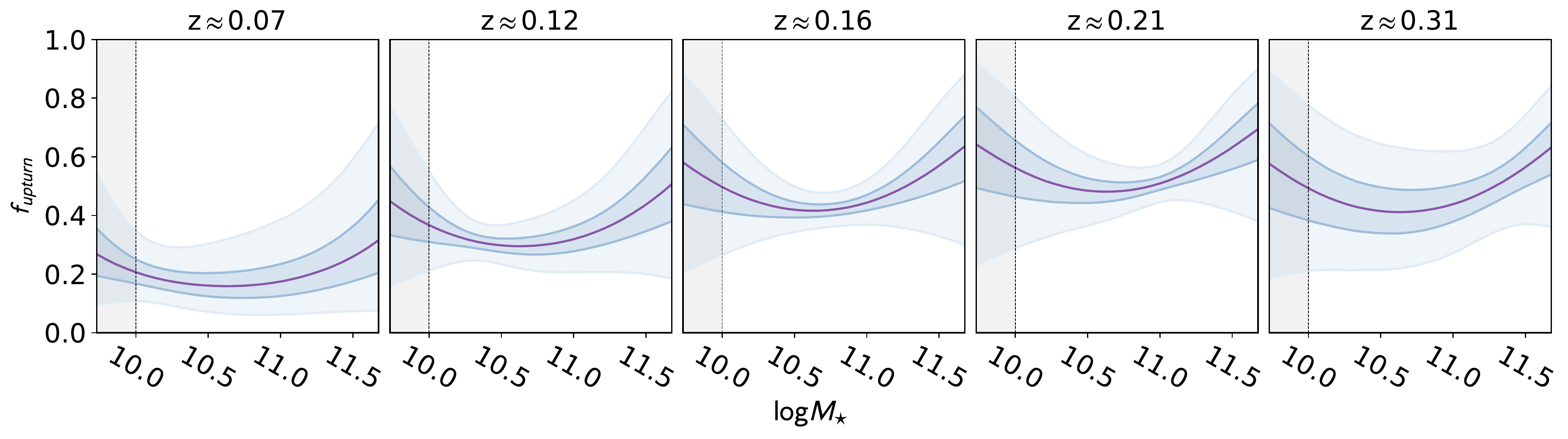}
    \caption{Fraction ($f_{\mathrm{upturn}}$) of UV bright RSGs that foster the UV upturn phenomenon according to their $\log M_{\star}$. The blue shaded areas depict 50 and 95 per cent probability intervals. A dashed line is displayed at $\log M_{\star} = 10.0$ to remind the reader of the small number of objects with $\log M_{\star} < 10.0$; such area is also tinted in light grey.}
    \label{fig:regress_mass_noemlines}
\end{figure*}

Fig. \ref{fig:regress_mass_noemlines} displays the results as a function of $\log M_{\star}$. This projection is shown for five slices of $z$ and the rough overall trend remains the same throughout $z$. The fraction seems to decrease for $\log M_{\star} \approx 10$ and, then, it seems to rise again for higher values of $\log M_{\star}$. This up-rising shift (the moment the slope changes) moves towards higher masses for increasing $z$: for $z \approx 0.12$ it happens at $\log M_{\star} \approx 10.5$, whereas for $z \approx 0.21$ at $\log M_{\star} \approx 11$. Nonetheless, for $\log M_{\star} \lessapprox 10.0$ the credible intervals become too wide; this is due to the very low number of objects with low mass as shown in Fig. \ref{fig:mass_cumulative}. To guide the reader, we added a dashed line at $\log M_{\star} = 10.0$ and tinted the area below it in light grey.

\begin{figure*}
    \centering
    \includegraphics[width=\linewidth]{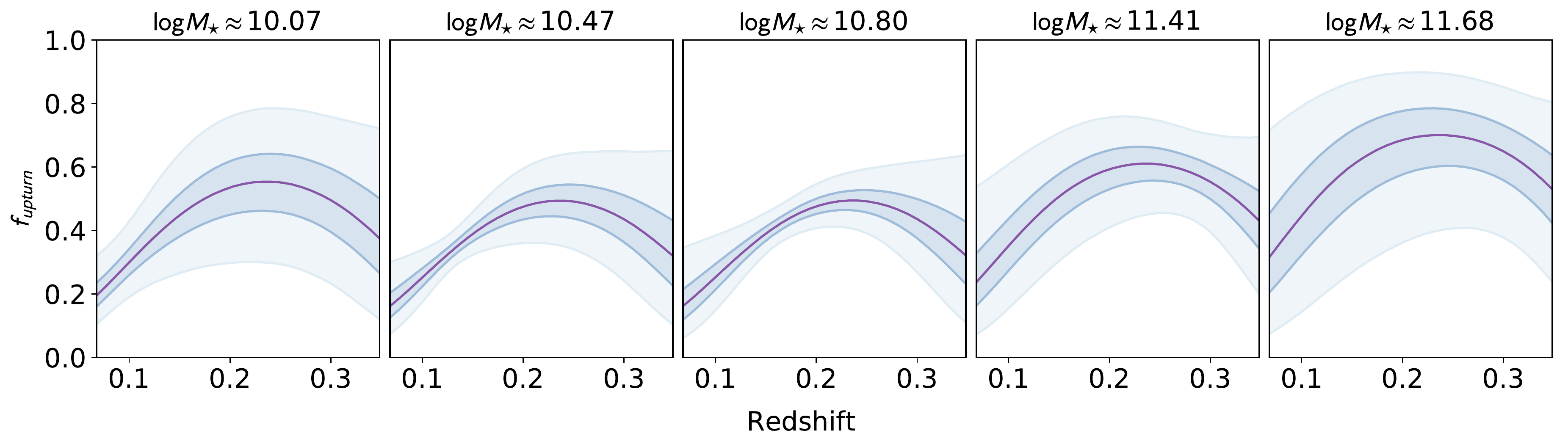}
    \caption{Fraction ($f_{\mathrm{upturn}}$) of UV bright RSGs that foster the UV upturn phenomenon according to their $z$. The blue shaded areas depict 50 and 95 per cent probability intervals.}
    \label{fig:regress_z_noemlines}
\end{figure*}

Fig. \ref{fig:regress_z_noemlines} displays the results for the regression in terms of $z$ for five slices of $\log M_{\star}$. Likewise, the overall trends remain the same for each slice displayed: the fraction of RSGs carrying UV upturn increase with $z\approx0.06$ to $z\approx0.25$, which is followed by what seems to be an in-fall; however, with such decrease, the credible intervals widen considerably, and one cannot safely affirm whether the probability decreases, increases, or plateaus.

\subsection{Results for stellar mass, redshift, and WHAN classes} \label{subsec:results02}

Similarly to Sec. \ref{subsec:results01}, this one features the model results for $\log M_{\star}$ and $z$, however stratified by emission-line classes according to the classes depicted in the WHAN diagram. 

\begin{figure*}
\includegraphics[width=\linewidth]{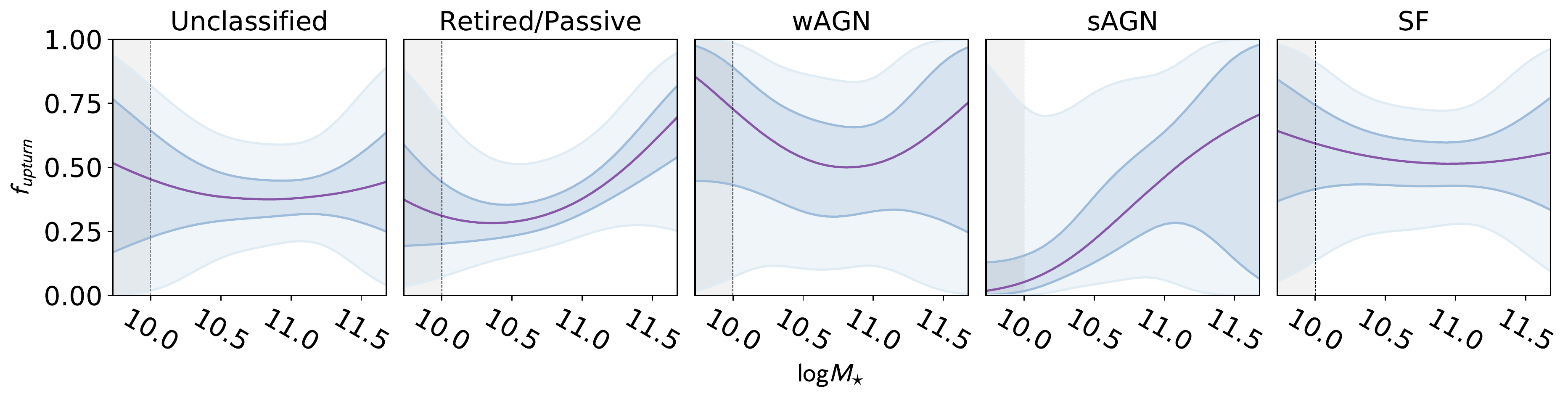}
\caption{Fraction ($f_{\mathrm{upturn}}$) of UV bright RSGs that foster the UV upturn phenomenon according to their $\log M_{\star}$, stratified by their emission-line classes. In this image, the $z$ slice is the median of our sample: $z\approx 0.21$. A dashed line is again displayed at $\log M_{\star} = 10.0$, as well as the grey area. The blue shaded areas depict 50 and 95 per cent probability intervals.}
\label{fig:regress_mass}
\end{figure*}

\begin{figure*}
    \includegraphics[width=\linewidth]{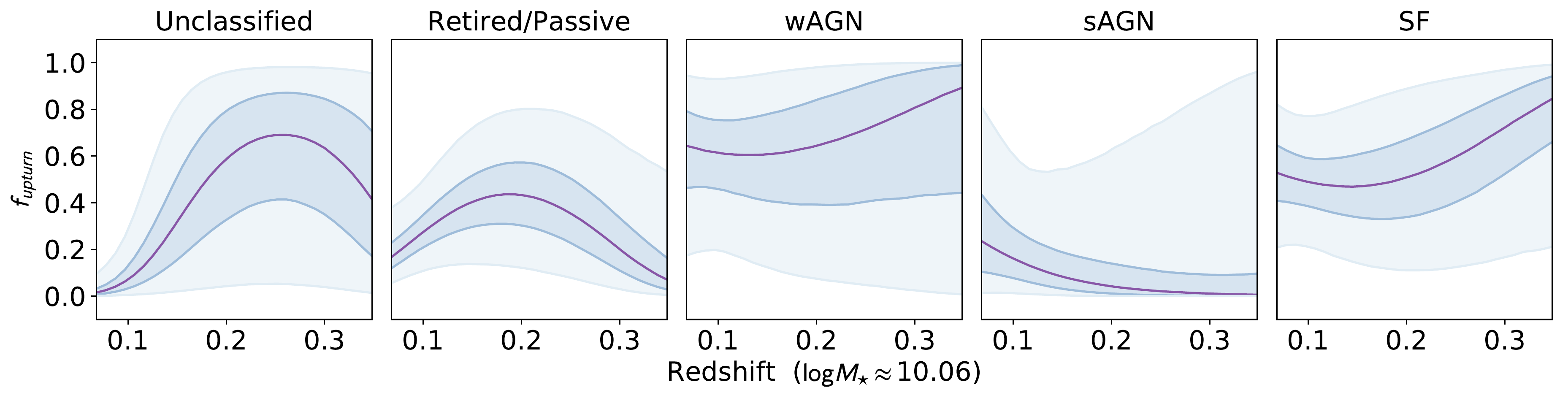}
    \includegraphics[width=\linewidth]{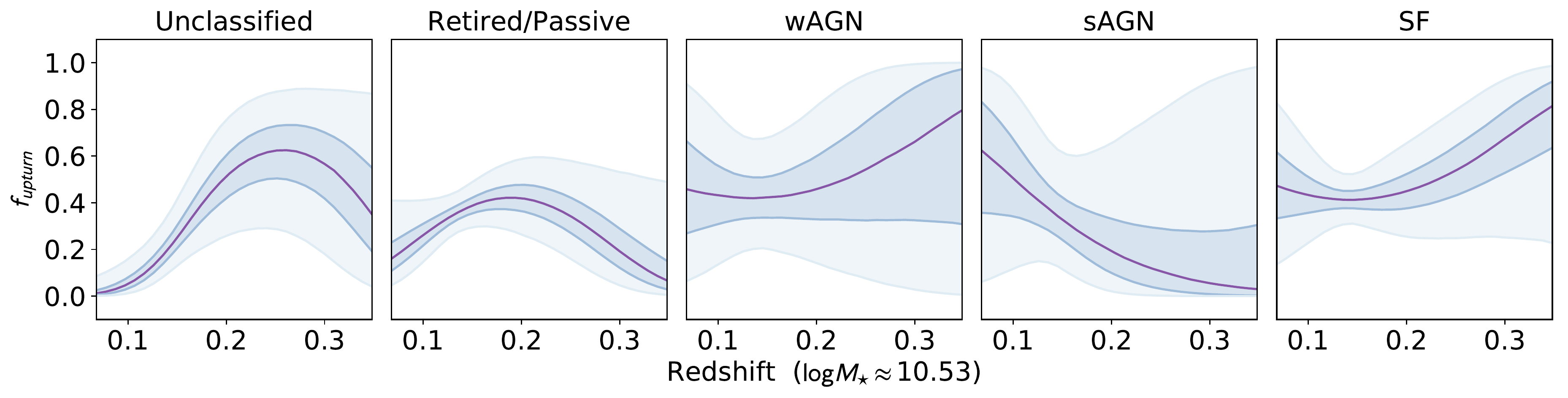}
    \includegraphics[width=\linewidth]{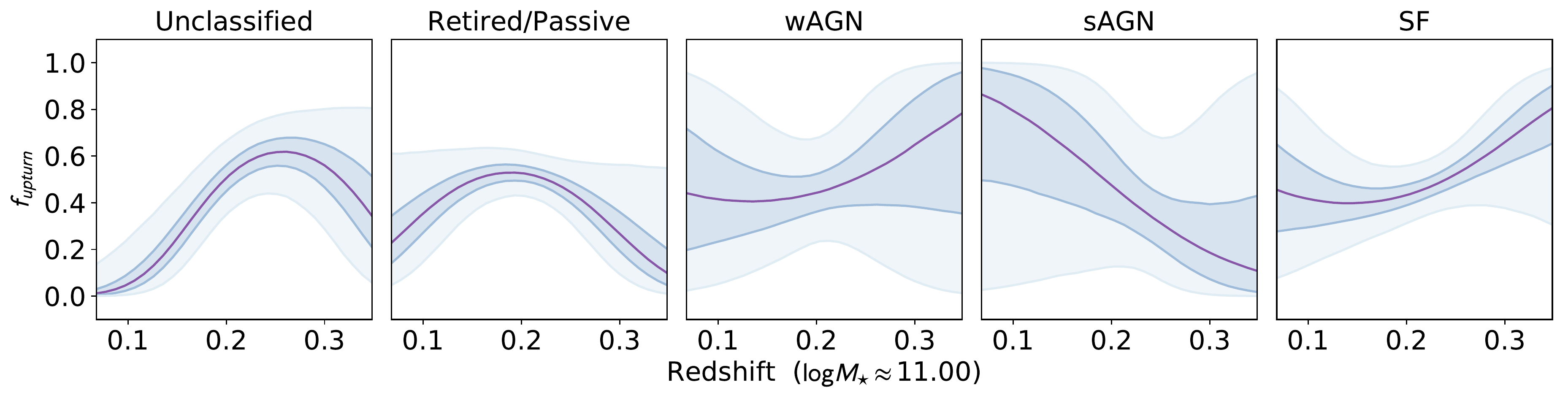}
    \includegraphics[width=\linewidth]{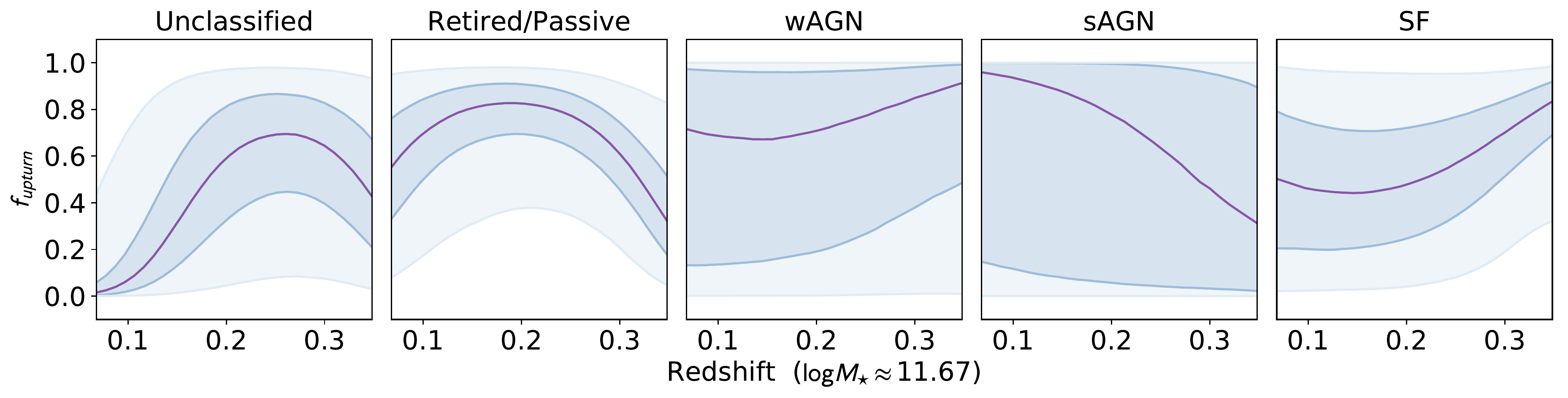}
\caption{Fraction ($f_{\mathrm{upturn}}$) of UV bright RSGs that foster the UV upturn phenomenon as a function of $z$, stratified by their emission-line classes. Projections of $f_{\mathrm{upturn}}$ are displayed for four slices of  $\log M_{\star}$, as shown in each corresponding x-axis. The blue shaded areas depict 50 and 95 per cent probability intervals.}
\label{fig:regress_z}
\end{figure*}

Fig. \ref{fig:regress_mass} illustrates the fraction of an UV upturn system within the UV bright RSG population, given $\log M_{\star}$ and their emission line classes. Likewise, Fig. \ref{fig:regress_z} displays the probability of an UV bright RSG to present the UV upturn phenomenon at a given $z$; the results are displayed according to their emission-line class, given four slices of $\log M_{\star}$.

\section{Discussion}\label{sec:discussion}

The strange UV emission from early-type systems seems to be caused by a myriad of competing phenomena, which can substantially vary from galaxy to galaxy. UV bright RSGs can nest a small fraction of newborn stars, but also evolved post-main-sequence stellar populations (e.g. post-AGB, HB stars), as well as rare stellar content (e.g. binary systems). Other potential players, such as AGN, also need to be taken into account when dealing with such intricate phenomenon.

Differently from \citet{Ali2018c} -- whom explored whether there was an evolution of the UV upturn by using selected systems located in four galaxy clusters in three distinct redshift bins (0.31, 0.55, 0.68) using HST archive data --, we made use of observations from the GAMA survey, considered the fraction of UV upturn systems among UV bright RSGs, and used systems in a continuous range of $z$ between 0.06 and 0.35--0.40. Our results suggest that, in fact, there is a dependency between the presence of UV upturn in RSGs with $\log M_{\star}$ and $z$ as shown in Figs. \ref{fig:regress_mass}, \ref{fig:regress_z}, and Sec. \ref{subsec:results01}.

It is important to disclose that boundary classification issues can also be impacted by large k-corrections -- specially in the UV, as $z$ increases -- as shown in Fig. \ref{fig:kcorr}. In other words, galaxies near the classification border are more likely to be misclassified. Hence, the current approach may benefit from a further exploitation of different sources of classification biases.

In what follows, we discuss the dependencies with $z$ and $\log M_{\star}$ stratified by WHAN class to better guide the reader along with the results presented in Figs. \ref{fig:regress_mass} and \ref{fig:regress_z}.

\subsection{Unclassified and R/P}
Quiescent systems are the classical the UV upturn bearers, which are represented these classes. Their stellar content is dominated by evolved stellar evolutionary phases. 

\paragraph*{Dependency with $z$:} the trend seen in Fig. \ref{fig:regress_z_noemlines} seems to be mainly influenced by unclassified, R/P, and SF systems (as shown in Fig. \ref{fig:regress_z}). Apparently quiescent galaxies tend to present an in-fall of the fraction of UV upturn among UV bright systems for $z \gtrapprox 0.2-0.25$ (equivalent lookback time of 2.5--3 Gyr). SF objects present a slight decreasing fraction in lower $z$, which rapidly changes, rising from $z \approx 0.15$ ($\approx$ 1.9 Gyr) until the $z$ limit of our sample.

\paragraph*{Dependency with $\log M_{\star}$:} while it is unclear whether there is a mass dependency for unclassified systems, R/Ps show a way more defined trend; the fraction of UV upturn systems rises with increasing $\log M_{\star}$. This indicates that the overall trend shown in Fig. \ref{fig:regress_mass_noemlines} seems to be driven mainly by R/P galaxies (Fig. \ref{fig:regress_mass}), since wAGN and sAGN show no reliable trend, and unclassified and SF systems do not seem to significantly change for increasing mass, which is further discussed in the following Secs. \ref{subsec:disc_wsAGN} and \ref{subsec:disc_sf}. Differently from the results by \citet{Yi2011}, our results point to a dependency of the incidence of UV upturn with $\log M_{\star}$ -- and potentially to velocity dispersion as discussed by \citet{burstein+88}, which remains to be further explored.

\subsection{wAGN \& sAGN} \label{subsec:disc_wsAGN}

In order to address both AGN classes displayed in the WHAN chart, one must consider the a few issues, such as the arbitrary boundaries, as well as the AGN unification theory \citep{Antonucci1993} and its controversies \citep[e.g.][for a comprehensive review on this issue]{Elitzur2006, Netzer2015}. 

In fact, if we take the BPT into account, there are no Seyferts in this sample (we refer the reader to Appendix \ref{sec:app_bpt}) -- hence, no `real' strong AGN -- classified as UV upturn objects. Moreover, the WHAN diagram for UV weak and upturn shown individually in Fig. \ref{fig:bptwhan_split} highlight the fact that the sAGN of the sample is very close to the surrounding boundaries (hence to the other classes), specially for the latter. The best assumption is that the sAGN of our sample are actually misclassified. The advantage of this result is that the criteria described by \citet{Yi2011} seems to be robust against contamination from \emph{bona fide} AGN. Moreover, it is important to keep in mind that most AGN inhabit galaxies in the green-valley \citep{Smolcic2009, Schawinski2010}, not to mention that Seyferts are predominantly nested by spirals and lenticulars \citep[see, for instance,][]{Hunt1999,OrbandeXivry2011}; hence, it is reasonable that our sample is clear of them. Yet, the potential contamination of green-valley systems, described in the end of Sec. \ref{subsec:finalsample}, can be linked to the detection of this small number of AGN.

The phenomenon attributed to LINER (which we understand as being a wAGN, as extensively discussed in \citealt{DeSouza2017}) galaxies has been a subject to debate in recent publications \citep[e.g][]{Sing2013, Belfiore2016}. It has been argued that many of these galaxies that have been linked to AGN activity are, in fact, behaving as such due to the presence of HOLMES \citep[e.g. post-AGB, HB stars][]{Sing2013}. In order to comprehend the potential link between LINERS and UV upturn, it is essential to understand which components of the galaxy are responsible for such emission \citep[as demonstrated by][]{Belfiore2016}.

In short, no strong affirmation can be made for wAGN or sAGN. Issues like classification boundaries, small numbers of (supposedly) AGN (which lead to wide credible intervals), the AGN unification theory strife, and the potential contamination of green-valley galaxies suppress strong physical meaning for them. Therefore, we do not carry on with the discussion for $\log M_{\star}$ and $z$ dependencies for these classes.

\subsection{Star-forming} \label{subsec:disc_sf}
Although SF activity may seem the least interesting ionisation source linked to UV upturn, several works have shown that this must not be overlooked \citep[e.g.][and references therein]{LopezCorredoira2018}. ETGs can be subject to rejuvenation due to several processes; for example, \citet{Bettoni2014} have shown that counter-rotating ETGs have recently undergone minor mergers have been shown to present non-negligible star-formation activity ($\approx 50$ per cent present strong far-UV emission).

\paragraph*{Dependency with $z$:} the fraction for SF increases with $z$, as expected given that the UV flux in such systems emanates from rapidly-evolving massive stars -- which are bluer, more massive, and less metallic as redshift increases \citep[e.g.][and references therein]{Madau1996, Vink2018}.

\paragraph*{Dependency with $\log M_{\star}$:} large credible intervals can be seen at all given $\log M_{\star}$. The overall results indicate that no trend can really be seen, also are similar to unclassified galaxies. Additionally, very massive galaxies tend to be quiescent, which indicates that SF systems in the high mass end seem to be misclassified.

\subsection{Remarks on previous works}

Exploring the evolution of any phenomenon within $z$ is not an easy task; the impact of selection biases permeate the entire analysis and it is often difficult to assess their impact. Previous works have attempted to probe whether the UV upturn evolves with increasing $z$, but each endeavour has had a particular approach. Most of these works tackle whether the strength of the FUV flux changes with $z$ in well-known elliptical systems that foster the UV upturn phenomenon \citep[approach adopted by][]{Brown1999, Brown2004, Ree2007, Ali2018b, Ali2018c}.

\citet{Brown1999} used a sample of bright elliptical galaxies and compared their UV emission to theoretical SED predictions from \citet{Tantalo1996}. This study was inconclusive and the author stated that there was no clear evidence of the UV flux evolving with $z$.

Later \citet{Ree2007} performed a detailed analysis on a sample of twelve BCGs observed by GALEX (a combination of DIS and MIS observations) from the Abell clusters and compared it to predictions associated with HB and post-AGB stars. They concluded that the FUV flux fades with increasing $z$, in accordance with the predictions of population synthesis models retrieved from \citeauthor{Yi1999} (\citeyear{Yi1999}).

\citeauthor{Ali2018b} (\citeyear{Ali2018b}, \citeyear{Ali2018c}), similarly to \citeauthor{Brown1999} (\citeyear{Brown1999}, \citeyear{Brown2004}), have also explored the evolution of the strength of the UV upturn among selected objects observed by HST. Their results indicate an up-rise of the strength of the UV upturn up to $z~\sim 0.55$ followed by an in-fall.

Our results contrast with those of previous works for many reasons. First and foremost, the concept behind the present study is not of verifying whether the FUV flux strength changes with $z$, but to check how the proportion of UV upturn systems varies in $\log M_{\star}$ and $z$ (and emission line characteristics). Additionally, the criteria used to select the sample is more supple, being very different from the aforementioned works. That lead us to embrace a sample of RSGs by making use of colour-colour cuts, instead of strictly well-known elliptical systems.

All in all, we understand that the results herein presented do not pose a conflict with previous studies; they rather serve as enrichment by looking at this phenomenon from a different standpoint.

\section{Summary \& conclusions} \label{sec:conclusions}

To  address the rate of occurrence of the UV upturn among UV bright red-sequence galaxies, we built a Bayesian logistic regression model, with the following parameters: stellar mass, redshift, and -- later also -- emission-line classes retrieved from the WHAN diagram. Galaxies that lack emission-line measurements were labelled as unclassified (which, as previously mentioned, are a mix of lineless passive systems and other galaxies with undetected emission-lines) and were also taken into account. Yet, it is the first time that a Bayesian hierarchical model is applied to scrutinise this phenomenon. The results herein presented are robust and are backed by a supplementary analysis with a volume-limited sub-sample which is available in Appendix \ref{sec:app_vollim}.

Our main results can be summarised as follows.

\begin{enumerate}[1.]
    \item Our results indicate that the photometric criteria proposed by \citet{Yi2011} is robust enough against \emph{bona fide} AGN contamination. On the other hand, the presence of SF systems is widely present (which correspond to, approximately, 26 and 32 per cent for UV upturn and UV weak systems, respectively). Also, the lack of intermediate regions between classes make the boundary areas questionable.
    
    \item From our overall exploratory analysis, the UV upturn systems heretofore investigated feature astrophysical characteristics that are consistent with previous works; they are brighter, redder in the optical, and more massive, when compared to their UV weak counterparts. 
    
    \item The results retrieved from our model suggest that the incidence of the UV upturn phenomenon strongly depends on redshift and stellar mass.
    
    \item By diluting the sample into emission-line classes, it is noticeable that:
    
    \begin{enumerate}[i)]
        \item the results for weak and strong AGN are very vague and no conclusions can be made for them;
        
        \item retired/passive objects (in the WHAN diagram) display the most clear-cut trends and results.
    \end{enumerate}
    
    \item Retired/passive systems present an up-rise in the fraction of UV upturn hosts for redshifts between 0.06 and 0.20--0.25, approximately. Then, an in-fall can be clearly seen.
    
    \item Retired/passive objects also show the strongest dependency with stellar mass, suggesting that the fraction of UV upturn hosts rise with increasing mass.
    
    \item Differently from former studies that focused their goal on determining whether the flux of the far-UV increases/decreases with increasing redshift, the present work aims at estimating the evolution of the \textit{rate} of occurrence of UV upturn among UV bright RSGs. Therefore, these results do not pose conflict to previous works, but serve as aggregation to the research on this subject.

    \item Similarly to previous studies, the results herein presented support the likelihood that the UV upturn is caused by a \emph{potpourri} of emitters, which remain to be properly quantified.
    
    \item It is important to highlight that the lack of correction for internal extinction is an non-negligible caveat, which can impact on their evolution status; highly obscured green-valley galaxies can pass as red-sequence.

\end{enumerate}

Results can potentially improve by using other criteria in order to classify systems with measurable UV emission. This can either be done by improving the boundary interface areas in the (NUV-$r$) $\times$ (FUV-NUV) colour-colour diagram \citep[similarly to what has been done in][]{Beck2016, DeSouza2017, Ucci2018}, or taking into account new observational data with narrow filters in the optical and UV \citep[e.g.][]{Benitez2014, MendesdeOliveira2019} -- or both.

Improvements in this analysis can be made also by investigating the fraction of UV upturn galaxies among the entire RSG population that are not necessarily UV bright.

Future follow-ups include a deeper investigation on the physical parameters of these objects via spectral energy distribution (SED) fitting with carefully chosen simple stellar population libraries (SSPs) for this purpose, which would enable us to disentangle even further the various drivers of the UV upturn phenomenon; as well as to better estimate the internal extinction thereby removing contaminating green-valley systems.

\section*{Acknowledgements}\label{sec:thanks}
MLLD acknowledges Coordenação de Aperfeiçoamento de Pessoal de Nível Superior - Brasil (CAPES) - Finance Code 001; and Conselho Nacional de Desenvolvimento Científico e Tecnológico - Brasil (CNPq) project 142294/2018-7. MLLD and PRTC thank A. Werle for the kind and fruitful discussions about this work. MLLD acknowledges D. Stoppacher for the kind review of the manuscript. MLLD and RSS also acknowledge Paçoquinha and Miuchinha for the support during the development of this work. PRTC acknowledges support from Fundação de Amparo à Pesquisa do Estado de São Paulo, FAPESP project 2018/05392-8, and Conselho Nacional de Desenvolvimento Científico e Tecnológico, CNPq project 310041/2018-0. RSS acknowledges the support from NASA under the Astrophysics Theory Program Grant 14-ATP14-0007 and U.S. DOE under contracts DE-FG02-97ER41041 (UNC) and DE-FG02-97ER41033 (TUNL)..
The authors also thank the anonymous referee for the substantial contribution to this work. 

We acknowledge the GAMA team for making their catalogues public and accessible.

The colours used in the figures herewith are colour-blind friendly and were retrieved from \url{www.ColorBrewer.org}. Credits: Cynthia A. Brewer, Geography, Pennsylvania State University.

This work benefited from the following collaborative platforms: \texttt{Overleaf}\footnote{{\url{http://overleaf.com/}}}, \texttt{Github}\footnote{{\url{https://github.com/}}}, \texttt{Slack}\footnote{{\url{https://slack.com}}}, and \texttt{SciServer}\footnote{{\url{http://www.sciserver.org/}}}.


\bibliographystyle{mnras}
\bibliography{references}

\appendix

\section{On data treatment details} \label{sec:app_challenges}

In this section we further explore details on data selection and treatment.

\subsection{On GALEX surveys} 

The GALEX satellite performed several surveys during its mission. We briefly recall its three main surveys and their overall characteristics, which are as follows.

\begin{enumerate}[1.]
    \item The All-sky Imaging Survey (AIS), with exposure time of 100 seconds, reaching AB magnitudes up to $\approx 20$ for FUV and $\approx 21$ for NUV, and covering 26,000 deg$^2$;
    \item The aforementioned MIS, with exposure time of 1500 seconds, reaching AB magnitudes up to $\approx 22.7$ for both UV bands, and covering 1,000 deg$^2$;
    \item The Deep-depth Imaging Survey (DIS), with whopping 30,000 seconds of exposure time enabling the detection of objects with AB magnitude up to $\approx 24.8$ and $24.4$ in FUV and NUV bands respectively, and covering modest 80 deg$^2$.
\end{enumerate}

For further details on these surveys, we refer the reader to \citet{Morrissey2007} and \citet{Bianchi2014}. The choice of MIS from GALEX DR6/7 was made considering the cross-match provided by the GAMA team \citep[we refer the reader to ][table 4]{Liske2015}. Ideally, the use of the DIS would be preferred for this study \citep[as vastly discussed by][]{Yi2008}; the depth reached by it would allow us better detect bright and very massive elliptical galaxies (such as BCGs) more accurately. Nevertheless, MIS was also able to catch some of the brightest elliptical galaxies observed by GALEX. Hence, in order to optimise the number of sources, the match made by the GAMA team was paramount; we believe it is the best `value for money' available for the purpose of this study, which is to investigate the \emph{fraction} of UV upturn systems among a larger sample of RSGs.

\subsection{On k-corrections \& internal extinction} \label{subsec:app_kcorr_extinct}

\subsubsection{K-corrections} 
\begin{figure*}
    \centering
    \includegraphics[width=\linewidth]{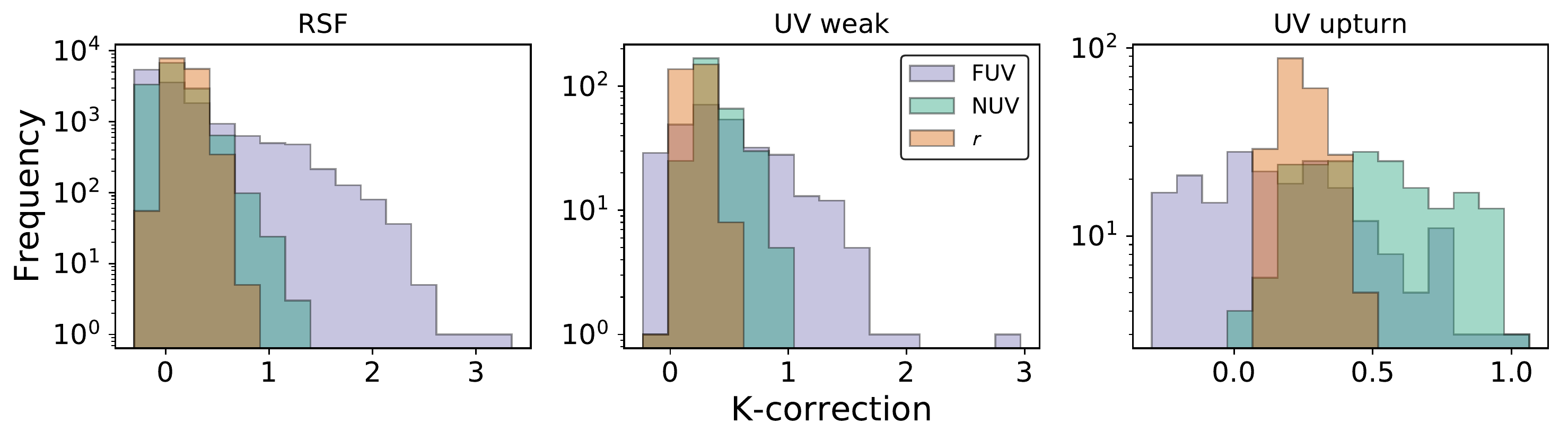}
    \caption{Distribution of k-corrections for the set of bands used to categorise the sample in UV classes: FUV in light purple, NUV in light blue, and $r$ in light red. Each panel depicts the different UV classes (i.e. RSF, UV weak and upturn, respectively) for the primary sample described in Sec. \ref{subsec:data_treat}.}
    \label{fig:kcorr}
\end{figure*}

We provide in Fig. \ref{fig:kcorr} the distributions of k-corrections applied for SDSS $r$-band and both GALEX bands, for the different UV classes.

\subsubsection{Internal extinction} 

We did not attempt to correct the colours in our sample for internal extinction. As such, it is possible that some UV upturn and weak galaxies with high extinction may be misclassified. To evaluate to what extend this can impact our analysis, we consider the following two factors:

\begin{enumerate}[i.]
    \item dust content;
    \item expected reddening at each band and their impact on the (FUV-NUV) and (NUV-$r$) colours.
\end{enumerate}

\citet{Werle2019} explore the efficiency of two different dust attenuation laws \citep{CCM1989, Calzetti2000} to simultaneously deal with SDSS and GALEX observations; from their prescription, we estimate the internal extinction in GALEX's UV bands and SDSS $r$-band as a function of $A_V$ can be expressed as:

\begin{subequations} \label{equation:fuv_nuv}
    \begin{flalign}
        A_{\rm FUV} &= 2.536 A_V;  \label{eq:afuv} \\ 
        A_{\rm NUV} &= 2.045 A_V;  \label{eq:anuv} \\
        A_{r}       &= 0.8695 A_V. \label{eq:ar}
    \end{flalign}
\end{subequations}

Werle et al. (in prep, private communication), investigate a sample of UV upturn systems in comparison to UV weak galaxies, with data combined from SDSS spectroscopy and GALEX photometry. Through SED fitting, they infer that the typical internal reddening of UV weak systems is about $A_V \sim 0.2$\,mag, resulting in the following extinction correction for the colours used to classify UV bright systems \citep{Yi2011}.

\begin{subequations} \label{eq:extinction}
    \begin{flalign}
    e_{\rm {\textsc{fuv-nuv}}} &= 0.491 A_V = 0.098; \label{eq:extinc_fuvnuv} \\
    e_{{\rm{\textsc{nuv}}}-r}  &= 1.176 A_V = 0.235; \label{eq:extinc_nuvr}   \\
    e_{{\rm{\textsc{fuv}}}-r}  &= 1.667 A_V = 0.333. \label{eq:extinc_fuvr}  
    \end{flalign}
\end{subequations}

\begin{figure}
    \centering
    \includegraphics[width=\linewidth]{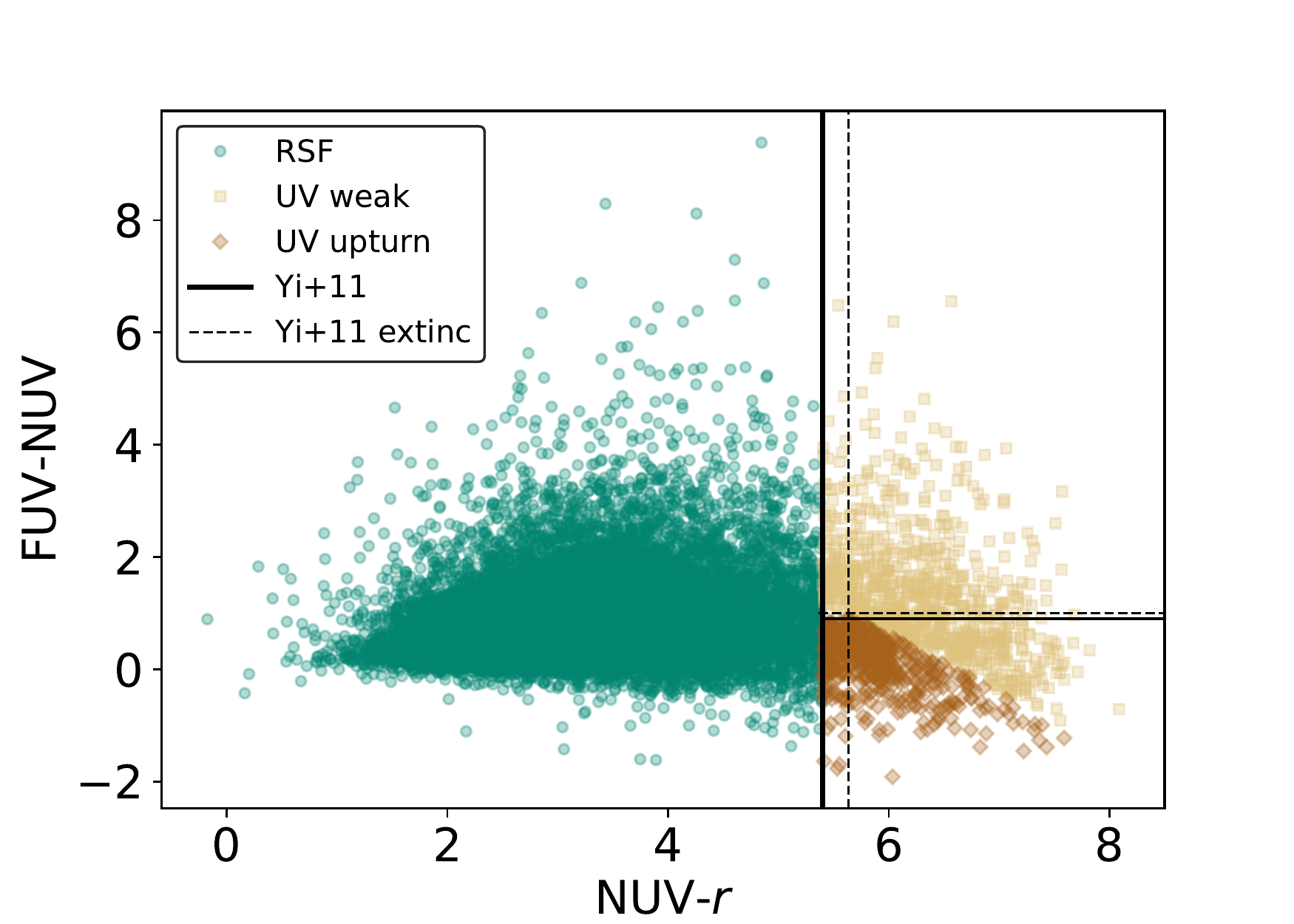}
    \caption{Similarly to Fig. \ref{fig:colourcolour_yi}, the above Fig. represents the colour-colour space with the UV classes construed by \citet{Yi2011}. Nonetheless, to visualise the impact of internal dust correction, we have displayed, their criteria by shifting the boundaries to reflect the extinction correction displayed in Eq. \ref{eq:extinction}. The original boundaries are represented by the continuous lines labelled as `Yi+11' and the extinction-corrected boundaries in dashed lines labelled as `Yi+11 extinct'. The objects between both lines (continuous and dashed) are the ones potentially misclassified. Additionally, the reddening suffered by the UV and UV-optical colours -- i.e. (FUV-NUV) and (NUV-$r$) -- are different, for the UV bands are more sensitive to dust and, thus, demand a larger extinction correction. In other words, both FUV and NUV bands need larger, however similar, corrections, which result in small changes in the (FUV-NUV) colour; whereas the $r$-band needs less corrections, leading to a larger colour correction for (NUV-$r$).} 
    \label{fig:app_colourcolouryi}
\end{figure}

The authors predict a lower dust content -- and therefore extinction correction -- for UV upturn systems. With these estimates on hand, we can infer that the overall correction for dust will induce some UV weak systems to migrate towards the UV upturn and, most importantly, to the RSF region; and a smaller proportion of UV upturn galaxies will migrate towards the RSF. To grasp the impact of such correction, Fig. \ref{fig:app_colourcolouryi} shows the necessary shift for the boundaries to reflect the internal extinction.

Considering all the three constrains in Eq. \ref{eq:extinction}, it is estimated that, from our \emph{primary} sample (displayed in Fig. \ref{fig:app_colourcolouryi}), the UV weak group would be depleted in $\sim$ 30 per cent -- with objects migrating to RSF and upturn -- and the UV upturn counterpart would be filled in $\sim$ 6 per cent -- loosing some objects to the RSF and receiving more from the UV weak migration.

\begin{figure*}
    \centering
    \includegraphics[width=\linewidth]{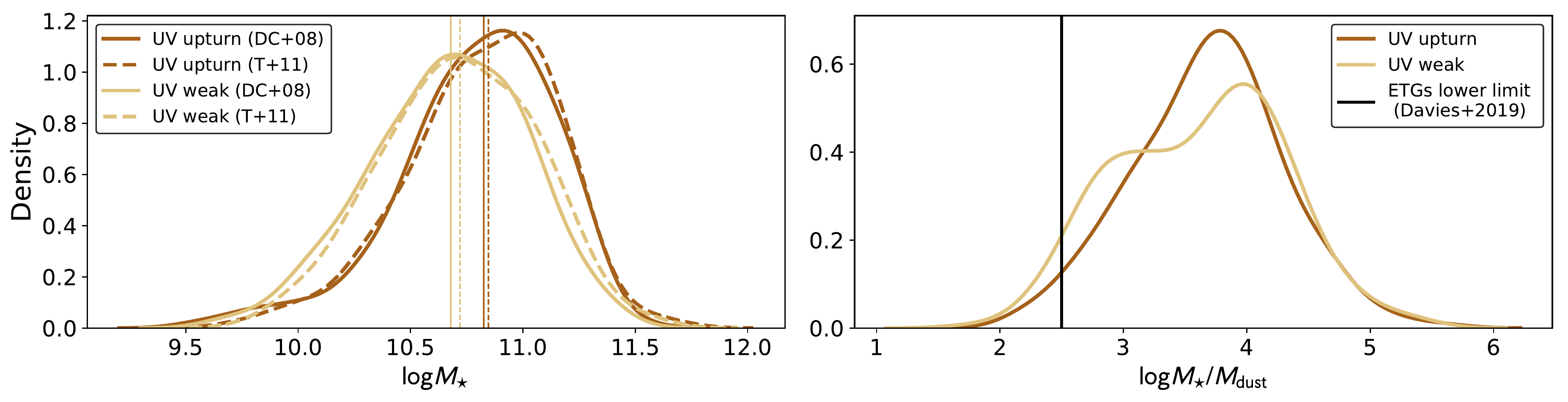}
    \caption{Left panel: for comparison, kernel density distributions for the resultant $\log M_{\star}$ estimated by \citet{Taylor2011} tagged as `T+11' and \citet{DaCunha2008} as `DC+08' for both UV weak and upturn systems; also their respective medians are displayed by the vertical lines with same coloured and shaped markers. The right panel shows the distribution of $\log M_{\star}/M_{\rm{dust}}$ measured by \texttt{MagPhys} with the black straight line as the lower limit at 2.5, established by \citet{Davies2019}.}
    \label{fig:dust}    
\end{figure*}

As a complementary step, we retrieved the dust parameters from \texttt{MagPhys} DMU from GAMA-DR3 \citep{DaCunha2008}, as described in Sec. \ref{subsec:data_selec}. Such parameters can be seen in Fig. \ref{fig:dust}, which has two panels: the left one displays the density distributions of $\log M_{\star}$ estimated by \citet{Taylor2011} -- which was used throughout this study -- and the mass determinations from \citet{DaCunha2008}, for both UV weak and upturn galaxies. Both estimates for $\log M_{\star}$ are in good agreement, allowing us to check the mass-to-dust ratio using the SED fitting results from \citet{DaCunha2008}, which takes us to the right panel; it displays the distribution of $\log  M_{\star}/M_{\rm{dust}}$ -- both retrieved from \texttt{MagPhys} -- with a vertical line indicating the lower limit of such parameter for ETGs according to \citet{Davies2019}.

The work of \citet{Davies2019} attempts to establish an overall relation between morphological classification of galaxies and their dust-to-stellar mass ratio, among other purposes. In the absence of external sources of dust (e.g. mergers), their work predicts a minimum value of $\log M_{\star}/M_{\rm{dust}}=2.5$ for E/S0 galaxies. By considering this lower threshold as a benchmark, we note that we have only 14 UV weak (4.7 per cent) and 5 upturn systems (2.4 per cent) which are heavily obscured by dust.

Given this issue, our overall remarks are:

\begin{enumerate}[i.]
    \item UV weak systems, when being properly corrected by internal extinction, tend to migrate towards the RSF region and some towards the UV upturn \emph{locus}. Likewise, some -- but not many -- of the UV upturn systems will migrate towards the RSF region (Werle et al., in prep., estimate smaller extinction corrections for UV upturn systems than those obtained for their UV weak counterparts). As a consequence, it is likely that our fractions are, in fact, underestimated.
    
    \item By making use of DustPedia predictions \citep{Davies2019}, we expect that only a small percentage of our UV bright RSGs will be, in fact, heavily impacted by dust-rich interlopers.
\end{enumerate}

\section{Complementary results from the BPT diagram} \label{sec:app_bpt}

Table \ref{table:bpt_uv} displays the cross-match between UV \citep{Yi2011} and BPT diagrams, including undetected objects -- tagged as unclassified. We consider a conservative composite area, being the one delimited by the lines of \citet{kewley2001} and \citet{Stasinska2006}, tinted in light grey. We also briefly explore the behaviour of the UV upturn galaxies categorised as R/P and unclassified according to the WHAN chart (see Tab. \ref{table:whan_uv}, last two rows and two columns). 

\begin{table} 
\caption{The following table displays the amount of galaxies of each UV class in the BPT diagram; it includes all the 11,647 with the required parameters successfully measured for the BPT plus the remaining objects, analogously to Table \ref{table:whan_uv}. Seyferts are displayed as `Sy', composites as `comp.' and unclassified as `unc.'.}
\label{table:bpt_uv}
\begin{center}
\begin{tabular}{lrrrrrr}
\multicolumn{7}{c}{\textbf{BPT classification}} \\
\hline
\hline
\textbf{UV class} & SF & Sy & LINER & comp. & unc. & \textbf{total} \\
\hline
RSF            & 9,223  & 37    & 197   & 1,975  & 2,402 & 13,834\\ 
\textbf{weak}  & \textbf{50} & \textbf{1} & \textbf{27} & \textbf{50} & \textbf{168} & \textbf{296} \\ 
\textbf{upturn} & \textbf{49}  & \textbf{0} & \textbf{20} & \textbf{27} & \textbf{114} & \textbf{210}\\
\end{tabular}
\end{center}
\end{table}

\begin{figure}
    \centering
    \includegraphics[width=\linewidth]{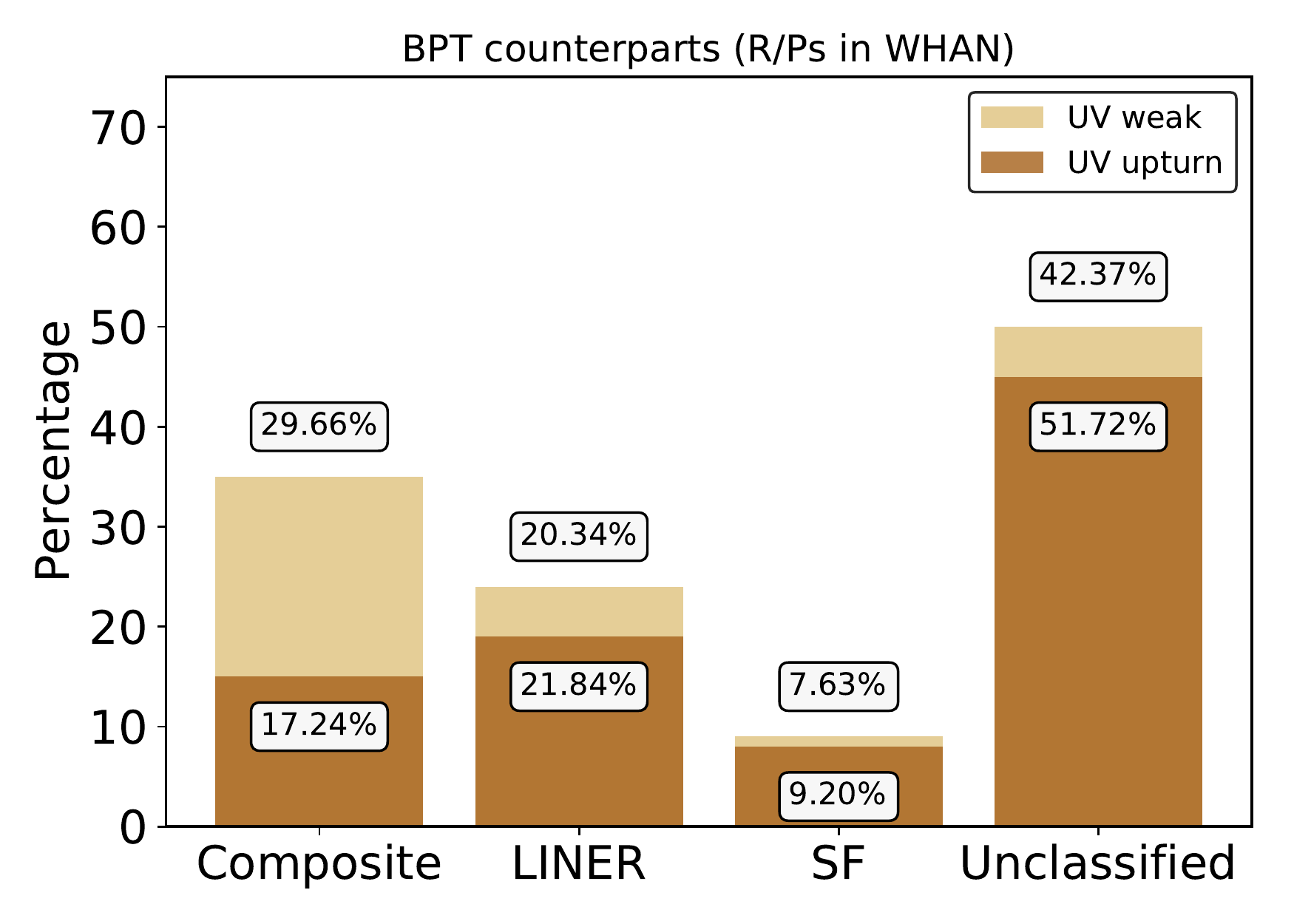}
    \caption{Bar-plots are displayed with the cross-classification of R/P galaxies in the WHAN diagram in terms of their counterparts in the BPT. UV weak and upturn systems are depicted in light and dark orange, respectively; and their precise percentages are displayed in the corresponding boxes over or under their respective bars. Yet, those tagged as `unclassified' in the WHAN diagram are also `unclassified' in the BPT and, therefore, were omitted. Also, no Seyferts were detected among UV bright R/P systems in our sample.}
    \label{fig:proportion_rp_and_na} 
\end{figure}

Fig. \ref{fig:proportion_rp_and_na} displays the classification in the BPT diagram for those objects that are known as R/P. Over 51 per cent of the WHAN R/P population is diagnosed as unclassified; LINER and composite account for almost 40 per cent of them; the remaining account for contaminating SF systems. This boundless feature has been discussed throughout the paper. The proportions of UV weak galaxies are also displayed, which enables us to compare the percentage of UV weak and upturn galaxies in each category. Overall, the proportions are somewhat similar, except for composite objects. This is probably due to the lack of certainty regarding the nature of the emitter responsible for these emission-lines, as well as the aforementioned boundary issues. Below we examine each BPT counterpart class labelled as R/P in the WHAN diagram, as displayed in Fig. \ref{fig:proportion_rp_and_na}.

\paragraph*{Unclassified} objects in the BPT are mostly a sum of R/P and unclassified systems from the WHAN diagram; these were discussed throughout the main Sections of this investigation.

\begin{figure} 
\includegraphics[width=\linewidth]{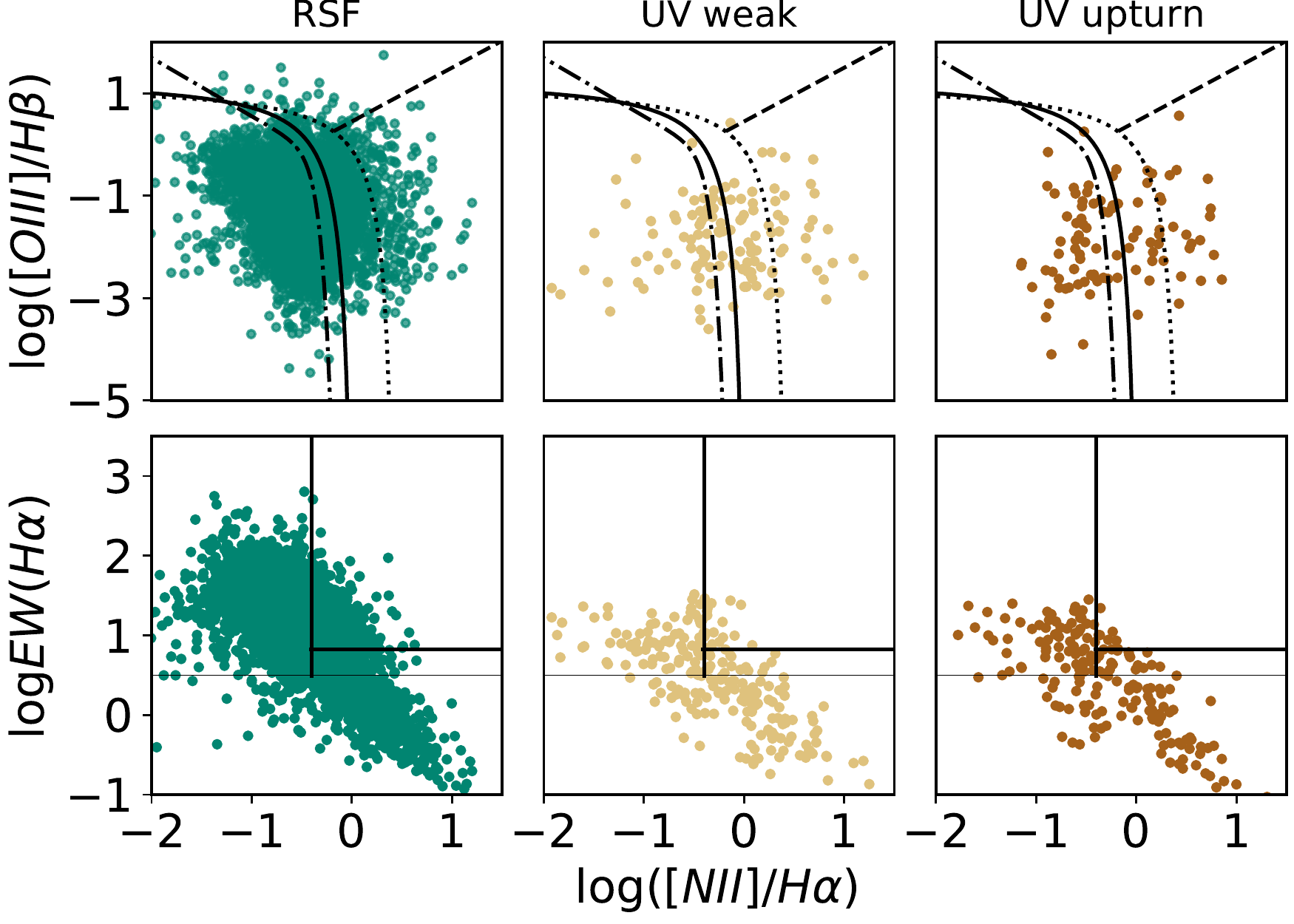}
\caption{BPT and WHAN stratified by UV class.}
\label{fig:bptwhan_split}
\end{figure}

\paragraph*{Seyferts} are absent in our UV upturn sample, and only one was detected as UV weak (the reader may notice the proximity of this object to the boundary line in Fig. \ref{fig:bptwhan_split}). As previously discussed, this is an important indicator that the method used by \citet{Yi2011} to select UV weak and upturn is robust against AGN. Moreover, Seyferts are mainly hosted by late-type systems \citep[see, for instance,][]{Hunt1999,OrbandeXivry2011}, thus it is reasonable that they are not caught within our selection method.

\paragraph*{LINERs} have also been discussed throughout the main Sections of this paper (we refer the reader to Sec. \ref{subsec:disc_wsAGN}).

\paragraph*{Composite} can vary a lot from galaxy to galaxy; in theory, the source of the emission-lines can be AGN -- including potential LINERs -- or star-forming, or all simultaneously.

\paragraph*{Star-forming} have also been discussed throughout the main Sections of this paper (we refer the reader to Sec. \ref{subsec:disc_sf}).

\section{Complementary results for a volume-limited sub-sample} \label{sec:app_vollim}

\begin{figure}
    \centering
    \includegraphics[width=\linewidth]{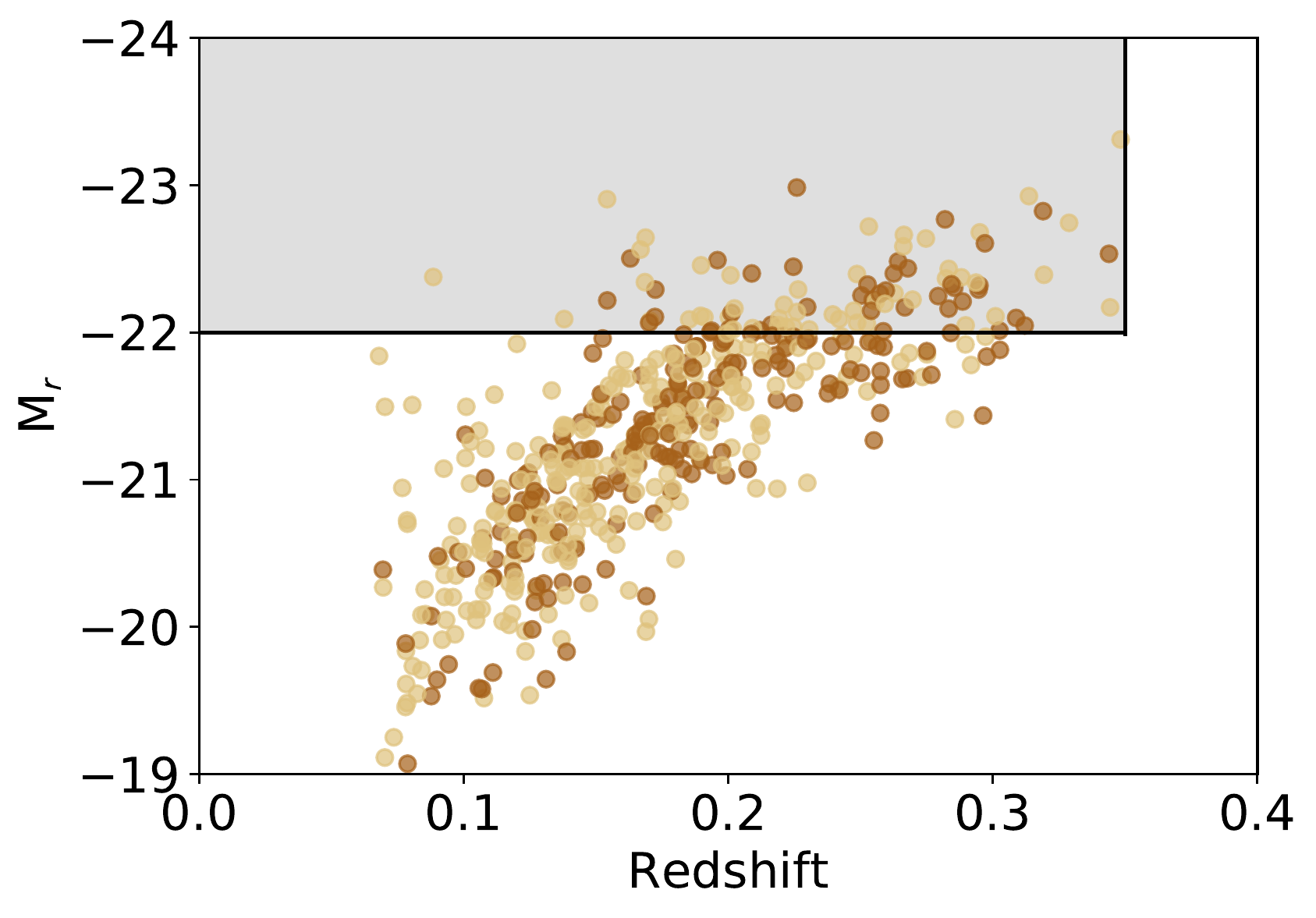}
    \caption{Analogously to Fig. \ref{fig:boxplots_mag} with the explicit volume cut filled in light grey: $M_r \leq -22$ and $0.06 \leq z \leq 0.35$. UV weak and upturn systems are coloured according to the same palette described in Fig. \ref{fig:colourcolour_yi}.}
    \label{fig:volume}
\end{figure}

We have selected a volume-limited sub-sample from the final sample described in Sec. \ref{subsec:finalsample}; that is: $M_r \leq -22$ and $0.06 \leq z \leq 0.35$, as illustrated in Fig. \ref{fig:volume}. This sub-sample dramatically reduces the number of objects into mere 91 RSGs, being 50 UV weak and 41 upturn systems (the original numbers can be found in Table \ref{table:whan_uv}.). Except for the sample, the model and parameters used are the same as those described in Sec. \ref{sec:methodology}.

\begin{figure*}
    \centering
    \includegraphics[width=\linewidth]{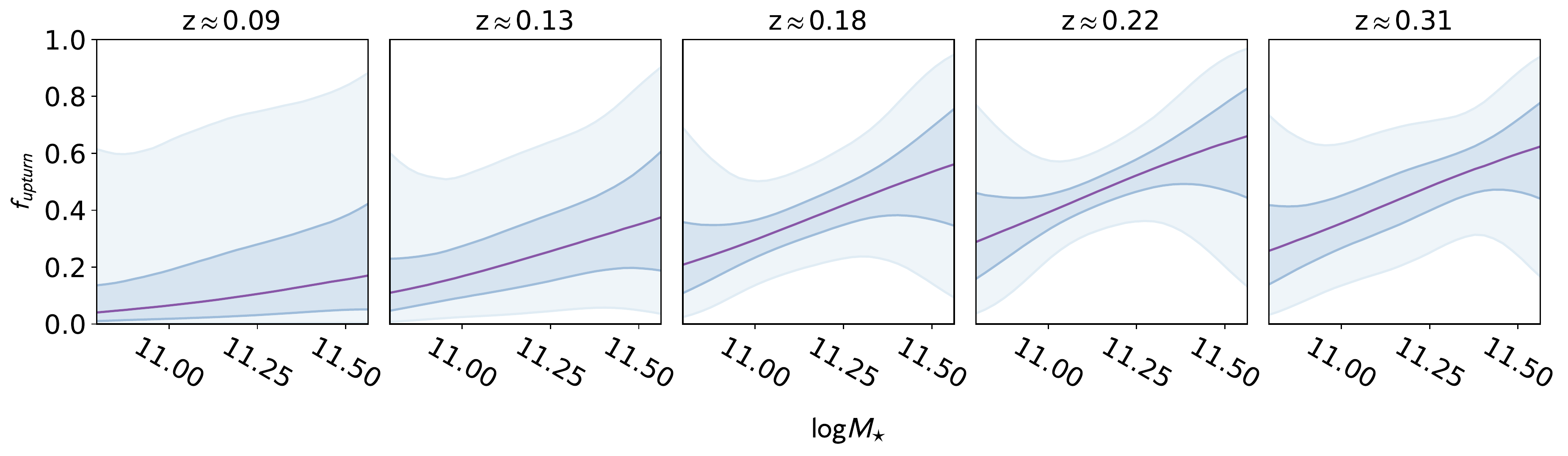}
    \includegraphics[width=\linewidth]{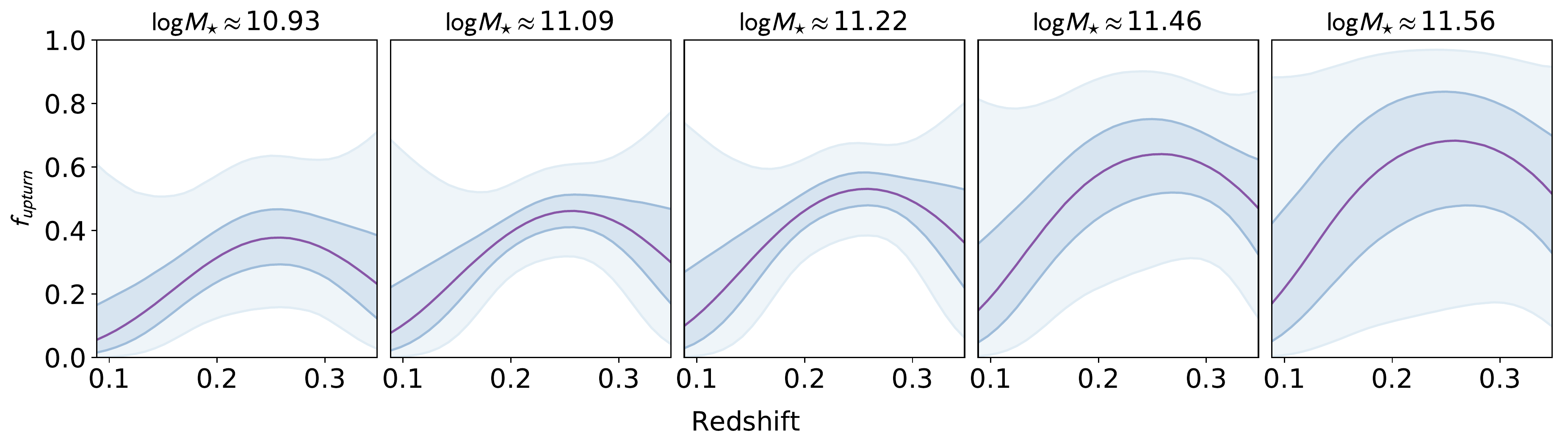}
    \caption{Analogously to Figs. \ref{fig:regress_mass_noemlines} and \ref{fig:regress_z_noemlines} respectively, with a smaller volume-limited sub-sample; i.e. $M_r \leq -22$ and $0.06 \leq z \leq 0.35$. As a consequence, the range of $\log M_{\star}$ is slimmer and concentrated towards higher values ($\log M_{\star} \gtrapprox 10.5$).}
    \label{fig:regress_VOLLIM}
\end{figure*}

Considering that the primary approach of this work is to retrieve the evolution of the fraction of UV upturn systems within UV bright RSGs, it is reasonable to assume that these systems suffer from the same biases with increasing $z$ and $\log M_{\star}$. Therefore, by choosing to consider the proportion of such galaxies, these biases are somewhat damped. This is confirmed by the complementary results with the volume-limited sub-sample analysis in this Section. Fig. \ref{fig:regress_VOLLIM} displays very similar trends as those of Figs. \ref{fig:regress_mass_noemlines} and \ref{fig:regress_z_noemlines} respectively; however, wider credible intervals appear as expected from the dramatically smaller number of objects when compared with the final sample.

Due to such small sub-sample, the results stratified by WHAN classes are not relevant for the volume-limited sub-sample herein used.

\bsp	
\label{lastpage}
\end{document}

%% file: model.tikz
\begin{adjustbox}{width=0.475\textwidth}
\begin{tikzpicture}[
  mtx/.style={
    matrix of math nodes,
    left delimiter={[},
    right delimiter={]}
  },
  subtxt/.style={below, font=\footnotesize}]

  \matrix[mtx] (Y) {%
    \eta_{11} & \ldots & \eta_{1k}\\
    \eta_{21} & \ldots & \eta_{2k}\\
    \vdots & \ddots &\vdots\\
    \eta_{n1} & \ldots & \eta_{nk}\\
  };
  \matrix[mtx, right=of Y] (X) {%
    1      & \log{M_{\star}}_{1} & \log{M_{\star}}_{1}^2&  z_{1} &  z_{1}^2 \\
    1      & \log{M_{\star}}_{2} &  \log{M_{\star}}_{2}^2 & z_{2} &  z_{2}^2 \\
    \vdots &         & \ddots &   & \vdots \\
    1      & \log{M_{\star}}_{n} & \ldots & & z_{n}^2 \\
  };
  \matrix[mtx, right=0.5cm of X] (beta) {%
    \beta_{11} & \ldots &  \beta_{1k}\\
     \beta_{21} & \ldots & \beta_{2k}\\
      \vdots & \ddots  &\vdots \\
       \beta_{j1} & \ldots & \beta_{jk}\\
  };

  \node at ($(Y.east)!0.5!(X.west)$) {$=$};%
  \node[above=2ex of X] (modmat) {$\eta_{i[k]} \equiv \log\left(\frac{p_{i[k]}}{1-p_{i[k]}}\right); \qquad \eta_{i[k]} = X_{ij} \beta_{jk}$ };
  \node[above=2ex of modmat] (mod)
    {$y_{i[k]} \sim \rm Bern\left(p_{i[k]}\right)$};
  \node[subtxt] at (Y.south) {$n\times k$};
  \node[subtxt, align=center] at (X.south) {$n\times j$};
  \node[subtxt] at (beta.south) {$j\times k$};
\end{tikzpicture}
\end{adjustbox}